\documentclass[useAMS,usenatbib,a4paper]{mn2e}
\title[Galaxy Zoo]{Galaxy Zoo: Morphologies derived from visual inspection of galaxies from the Sloan Digital Sky Survey\thanks{This publication has been made possible by the participation of more than 100,000 volunteers in the Galaxy Zoo project.}}
\author[Lintott et al.]{
\parbox[t]{16cm}{Chris J. Lintott$^{1}$\thanks{Email: cjl@astro.ox.ac.uk}, Kevin Schawinski$^{1}$\thanks{Email: kevins@astro.ox.ac.uk}, An\v{z}e Slosar$^{1,2}$, Kate Land$^{1}$, Steven Bamford$^{3}$,  Daniel Thomas$^{3}$,  M. Jordan Raddick$^{4}$,  Robert C. Nichol$^{3}$, Alex Szalay$^{4}$, Dan Andreescu$^{5}$, Phil Murray$^{6}$, Jan van den Berg$^{4}$\\}\\ 
$^{1}$Oxford Astrophysics, Denys Wilkinson Building, Keble Road, Oxford, OX1 3RH, UK\\
$^{2}$Berkeley Centre for Cosmological Physics, Lawrence Berkeley National Laboratory and Physics Department, Berkeley, CA 94720\\
$^{3}$Institute of Cosmology and Gravitation, University of Portsmouth, Mercantile House, Hampshire Terrace, Portsmouth, PO1 2EG, UK\\
$^{4}$Department of Physics and Astronomy, Johns Hopkins University, 3400 N. Charles St., Baltimore, MD 21218, USA\\
$^{5}$LinkLab, 4506 Graystone Ave., Bronx, NY 10471, USA\\
$^{6}$Fingerprint Digital Media, 9 Victoria Close, Newtownards, Co. Down, Northern Ireland, BT23 7GY, UK\\
}

\usepackage[dvips]{graphicx}
\usepackage{setspace}

\begin{document}

\date{March 2008}
\pagerange{\pageref{firstpage}--\pageref{lastpage}} \pubyear{2008}

\maketitle

\label{firstpage}

\begin{abstract}
In order to understand the formation and subsequent evolution of galaxies one must first distinguish between the two main morphological classes of massive systems: spirals and early-type systems. This paper introduces a project, Galaxy Zoo, which provides visual morphological classifications for nearly one million galaxies, extracted from the Sloan Digital Sky Survey (SDSS). This achievement was made possible by inviting the general public to visually inspect and classify these galaxies via the internet. The project has obtained more than $4 \times 10^7$ individual classifications made by $\sim 10^5$ participants. We discuss the motivation and strategy for this project, and detail how the classifications were performed and processed. We find that Galaxy Zoo results are consistent with those for subsets of SDSS galaxies classified by professional astronomers, thus demonstrating that our data provides a robust morphological catalogue. Obtaining morphologies by direct visual inspection avoids introducing biases associated with proxies for morphology such as colour, concentration or structual parameters. In addition, this catalogue can be used to directly compare SDSS morphologies with older data sets. The colour--magnitude diagrams for each morphological class are shown, and we illustrate how these distributions differ from those inferred using colour alone as a proxy for morphology.

\end{abstract}

\begin{keywords}
methods: data analysis, galaxies: general, galaxies: spiral, galaxies: elliptical and lenticular
\end{keywords}

\section{Introduction}

Dividing galaxies into categories based on their morphology, or shapes, has been standard practice since it was first systematically applied by \citet{Hubble}. It is perhaps surprising that sorting galaxies into categories which are suggested solely by their morphology produces classifications which broadly correlate with other, physical parameters such as the star formation rate or gas fraction. The fundamental distinction drawn is between galaxies with spiral arms and early-type systems\footnote{For the purposes of this paper we use the term `elliptical' rather than `early-type' as this is the description used on the Galaxy Zoo site. However, the term should be understood as including both elliptical and lenticular systems.}. For most of the twentieth century, catalogues of classified galaxies were compiled by individuals or small teams of astronomers (e.g. \citealp{Sandage, deVaucouleurs}). With the advent of modern surveys (such as the Sloan Digitial Sky Survey or SDSS, see Section \ref{sec:SDSS}) containing many hundreds of thousands of galaxies this approach was no longer practical. 

Anticipating the problem these surveys would cause, \citet{Lahav} compared classifications from a set of experts who considered a sample of just over 800 galaxies. Their motivation was to create a training set for neural networks, with the aim of automating the classification process. While such methods have indeed been developed \citep{Ball}, modern studies (e.g. \citealt{Bernardi, Lintott}) still separate early-type galaxies from spirals in large data sets by using proxies for morphology rather than by directly determining morphology itself. Typically, selection criteria based on galaxy properties such as colour, concentration index, spectral features, surface brightness profile, structural parameters or some combination of these are used (e.g. \citealp{Abraham, Conselice, Kauffmann, Scarlata, Strateva}). However, the use of each of these critera results in an unknown and potentially unquantifiable bias in the resulting sample of galaxies. In other words, although morphological labels are often used for the resulting catalogues, each of these criteria  produces a sample different from that obtained by true morphological selection. Comparing results from samples selected using different morphological proxies can therefore be misleading. 

Avoiding such confusion between categories is inherently desirable, but is of particular importance - to give just one example - for studies which seek to understand the influence of star formation on the larger-scale process of galaxy formation. The colour of a galaxy is often used as a proxy for morphology, but is also a direct consequence of and therefore depends on the star-formation history of the galaxy being studied. By directly classifying objects according to their morphology, the catalogue is sorted according to their dynamics; spirals are rotating, whereas ellipticals are tri-axial \citep{Binney}. Other information such as colours, the presence or absence of emission lines, or the galaxy profiles can then be used to investigate the properties of the classified galaxies, rather than being used in the classification itself. 

Several subsets of the SDSS have been classified by professional astronomers. \citet{Fukugita} recently compiled a catalogue of early-type objects by the visual inspection of $\sim 2500$ galaxies in the SDSS by three expert classifiers. This is an order of magnitude smaller than the sample used by \citet{Schawinski} for their study of AGN feedback in early-type galaxies. Their sample, (MOSES: MOrphologically Selected Ellipticals in SDSS), was obtained by carrying out manual inspection of all objects in the SDSS DR4 spectroscopic sample with redshift $0.05<z<0.10$ and r-band magnitude $r<16.8$. The resulting sample consists of 48,023 galaxies, or approximately 5\% of the complete SDSS galaxy sample (see below). This sample was then inspected to identify galaxies with an elliptical morphology. The importance of such a morphology-driven classification can be seen from the comparision of MOSES ellipticals with those selected by Bernardi et al (2003). Of the ellipticals selected by Bernardi et al., 5\% show emission lines indicative of star-forming activity compared to 18\% of the MOSES sample. The sample selected by morphology alone includes a set of star-forming galaxies that are excluded from samples selected by other methods.

Despite the desirability of pure morphological classification, the samples provided by SDSS and other modern surveys are simply too large for astronomers to visually inspect the entire catalogue. Furthermore, without multiple independent classifications of the same galaxy, it is difficult to establish how much confidence can be placed in the classifier or classifiers. Ideally, large numbers of independent classifications would be made for each galaxy in the sample, allowing the errors to be quantified. 

In this paper we present the results of an attempt to solve this problem by inviting large numbers of people to classify galaxies over the internet. This solution - known as `crowdsourcing' or `citizen science' -  had been successfully employed by projects such as Stardust@Home \citep{Westphal,Mendez}. This project was a search for interstellar dust particles in the sample collected by the \emph{Stardust} spacecraft from Comet Wild-2, with the initial selection of samples for further analysis being made by visual inspection. Galaxy Zoo involves an order of magnitude more participants than its predecesors, and is the first attempt to apply these techniques to astrophysical problems. Visual inspection is also an excellent method for serendipitous discovery of the unusual in any data set, and the more unusual objects discovered by Galaxy Zoo classifiers will be discussed in a series of future papers.

\subsection{The Sloan Digital Sky Survey}
\label{sec:SDSS}
The galaxies for this project were drawn from the Sloan Digital Sky Survey \citep{York}. The SDSS is a survey of a large part of the northern sky providing photometry in five filters; $u$, $g$, $r$, $i$ and $z$ \citep{Fukugita96}, covering approximately 26\% of the entire sky. We use the latest available data, contained in Data Release 6 (DR6; \citet{DR6}). The SDSS spectroscopic target selection algorithm \citep{Strauss} produces the Main Galaxy Sample, which includes all extended objects with Petrosian magnitude $r<17.77$ \citep{Petrosian}. All objects in this sample which the SDSS photometric pipeline \citep{Lupton} identified as a galaxy were included in the Galaxy Zoo database, regardless of whether or not such spectra have been obtained to date. This list included a total of 738,175 galaxies drawn from the SDSS main galaxy catalogue. In addition, objects which were not in the spectroscopic catalogue but which had already been observed and as a result classified as a galaxy by the SDSS spectroscopic pipeline were added to our list. This secondary selection comprised 155,037 objects drawn from both the main and luminous red galaxy SDSS catalogues. In all, 893,212 objects were included in our sample. It is reasonable to assume that the accuracy of classification of a galaxy will depend on factors which including the apparent size of the system and its surface brightness. However, as the biases were unquantified before the study was completed, no cuts on this inclusive sample were imposed. 

\section{Galaxy Zoo}

The data for this project was collected via a website\footnote{www.galaxyzoo.org}. In order to minimize the degree of knowledge needed by the volunteers, users of the site were not required to distinguish between elliptical and lenticular galaxies, or between different classes of spirals (Sa, Sb etc). Visitors to the site were asked to read a brief tutorial giving examples of each class of galaxy, and then to correctly identify a set of `standard' galaxies. These standard systems were selected from the SDSS and classified by team members; those with a low degree of agreement were rejected. Those who correctly classified more than 11 of the 15 standards were allowed to proceed to the main part of the site. The bar to entry was kept deliberately low in order to attract as many classifiers to the site as possible. 

\begin{figure}
\includegraphics[width=0.35\textwidth,angle=270]{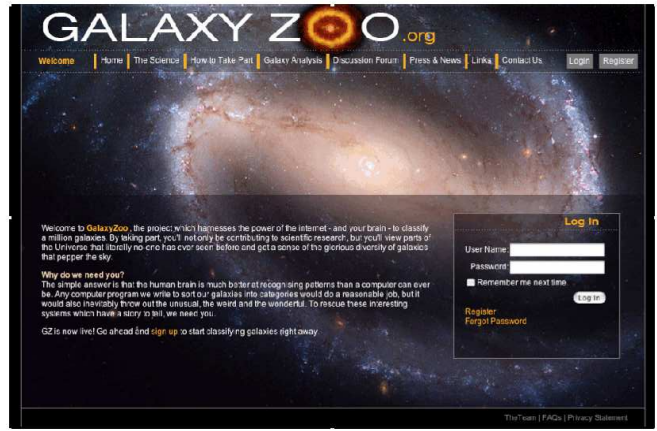}
\includegraphics[width=0.35\textwidth,angle=270]{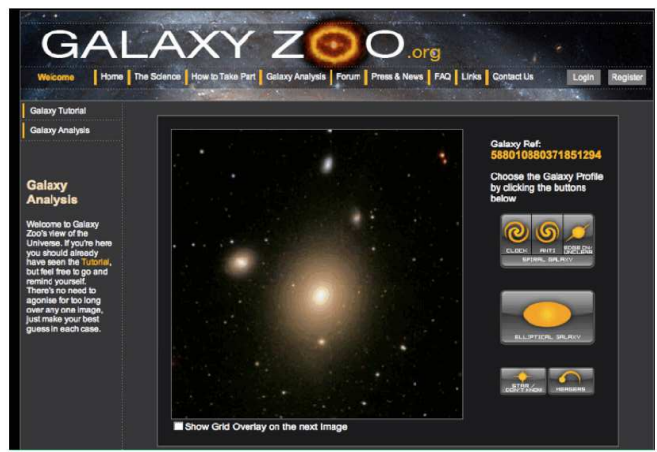}
\caption{Front page (top) and main analysis page (bottom) from the Galaxy Zoo website.}\label{fig:screenshots}
\end{figure}

The front page of the site and the main classification page are shown in Figure \ref{fig:screenshots}. SDSS images are shown to volunteers using the ImgCutout web service \citep{Nieto-Santisteban} on the SDSS website (Szalay et al. 2002). The service displays a JPEG cutout image of an area of sky, centered on a galaxy randomly chosen from the sample database, with an image scale of 0.024$R_p$ arcseconds per pixel where $R_p$ is the Petrosian radius for the galaxy. These images are colour composites of the three middle filters available in SDSS ($g$,$r$ and $i$). Details of the conversion to colour images are given in Lupton et al. 2004. Traditional morphological classifications have used single-band images in order to avoid confusion between morphology and colour. That said, these colour images are particularly suitable for visual classification. In particular, they possess the large dynamic range necessary for the identification of faint features, and have a unique mapping between physical and display colours. The effect of this choice on the data is discussed in section \ref{sec:colors}.  

In addition to sorting galaxies according to their morphology, the website asked classifiers to further divide galaxies they identified as spiral into three sub-categories according to the direction of their spiral arms (clockwise/anticlockwise/edge-on). This is a reliable indicator of the sense of the rotation of the galaxy \citep{Pasha}. The motivation for this part of the study is twofold. Firstly, we aim to investigate the evidence for a preferred handedness of spiral galaxies reported in the SDSS by Longo (2007). This result conflicts with an earlier paper by \citet{Sugai}, who did not find such a result in a different (but comparably sized) dataset. Longo's work was based on a sample of 2817 spirals selected by eye from galaxies in the SDSS with a redshift less than 0.04 and a magnitude of $g<17$. We have been able to extend his analysis to a sample which contains a factor of ten more galaxies, and the results are presented in a companion paper (Land et al., 2008). Secondly, it will also prove possible to use our results to calculate the two point correlation function for rotating spirals, an interesting new constraint on models of galaxy formation (Slosar et al. 2008).

Each object extracted from the SDSS database was thus classified as belonging to one of six categories: Spiral (Clockwise rotation), Spiral (Anticlockwise rotation), Spiral (Edge-on/rotation unclear), Elliptical, Merger, or Star/Don't Know. The symbols used for this classification are shown in Table \ref{tab:buttons}. In order to keep the task as simple as possible, no further distinction was made between barred and unbarred spiral systems, for example. Once a classification is chosen, then the image of the next galaxy is automatically displayed. 

As the Galaxy Zoo website gathers data, these are stored into a live
Structured Query Language (SQL) database. For each entry we store the timestamp, user
identification, galaxy identification and the classification chosen by
the user. Classifications by unregistred visitors are discarded and the user requested to register and complete the tutorial described above. For the analysis, this database may be downloaded and processed through the pipeline described below.

\begin{table}
\begin{tabular*}{0.45\textwidth}{@{\extracolsep{\fill}}ccl}
\hline
Class & Button & Description \\
\hline
1 & \raisebox{-0.5ex}{\includegraphics[width=0.5cm]{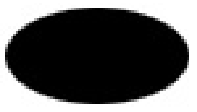}}
& Elliptical galaxy \\
2 & \raisebox{-2.0ex}{\includegraphics[width=0.5cm]{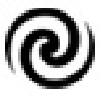}}
  & Clockwise/Z-wise spiral galaxy \\
3 & \raisebox{-2.0ex}{\includegraphics[width=0.5cm]{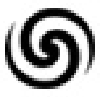}}
  & Anti-clockwise/S-wise spiral galaxy \\
4 & \raisebox{-2.0ex}{\includegraphics[width=0.5cm]{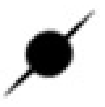}}
  & Spiral galaxy other (eg. edge on)\\
5 & \raisebox{-2ex}{\includegraphics[width=0.5cm]{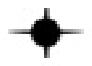}}
& Star or Don't Know (eg. artefact) \\
6 & \raisebox{-2ex}{\includegraphics[width=0.5cm]{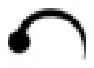}}
& Merger \\
\hline
\end{tabular*}
\caption{Galaxy Zoo classification categories showing schematic symbols as used on the site.}\label{tab:buttons}
\end{table}

Although some classifiers will inevitably know (or will learn from experience during the project) that spiral galaxies tend to be bluer than elliptical galaxies, the tutorial stressed that objects should be classified according to their morphology alone. No mention was made of the colour-morphology relation. In order to quantify the effect of colour on our results, a selection of monochrome images was introduced to the sample, and the results are discussed in Section \ref{sec:biasresults}. The data discussed in this paper represent the final results from the first stage of the project; Galaxy Zoo 2, which will ask for more detailed classifications, will follow shortly. 

\section{Producing a catalogue}

\begin{figure}
\includegraphics[width=0.45\textwidth]{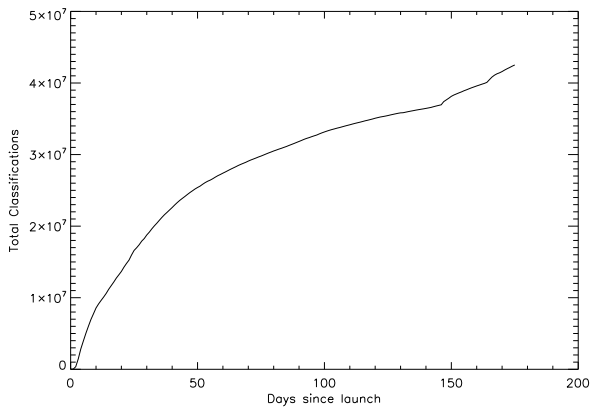}
\caption{Cumulative classifications collected by the Galaxy Zoo site. The sudden increase visible at $\sim 145$ and $\sim 160$ days correspond to email newsletters being sent out to those registered with the site. These led to a sustained increase in the rate of classification. Following day 140, data collected contributed to the bias study described in Section \ref{sec:biasresults}}\label{fig:class}
\end{figure}

\begin{figure}
\includegraphics[width=0.45\textwidth]{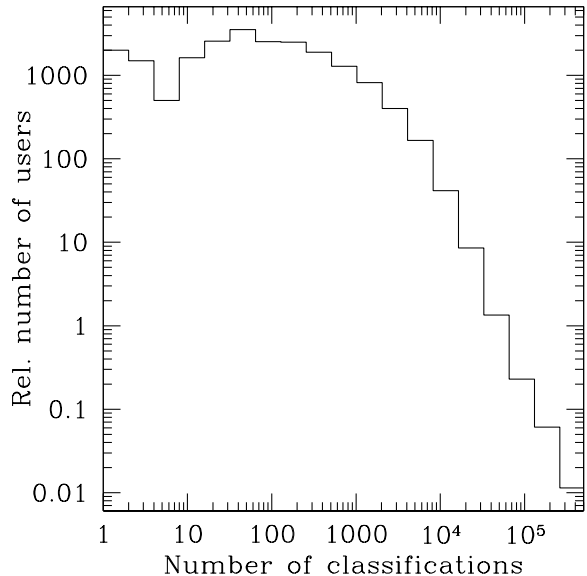}
\caption{The distribution of classifications among users. A small number have completed more than 100,000 classifications each, while the peak of the distribution is $\sim 30$ classifications per user.}\label{fig:class2}
\end{figure}

The website was successful in attracting large numbers of classifiers and classifications, as shown in Figures \ref{fig:class} and \ref{fig:class2}. Each galaxy in our sample was thus viewed and classified multiple times, with a mean of $\sim$38 classifications per galaxy. A variety of strategies are available to convert from these raw classifications to a final catalogue. In this section we compare the results from several different strategies. 

The first step in data reduction involves removing obviously bogus classifications. A small number of users seem to have recorded a number of these
classifications, either using some sort of automated
mechanism or due to some unknown problem with their browser. They are easy to discern by the fact that they have multiple classifications for a small number of galaxies. We find all users which have
classified two or more galaxies more than five times each. This is
extremely unlikely by Poisson distribution and hence \textit{all} data points
from such users are discarded. There are 36 such potentially malicious
users, amounting to less than 0.05\% of the total number of participants. Furthermore, in order to account for accidental double clicks, if a user has classified
the same galaxy more than once, we take into account only the first
classification from each user. This latter stage ensures that no single user can unduly influence the classification assigned to a single galaxy. The two steps of this cleaning process together remove about 4\% of our data set.

In the next step we create the so called \emph{combined spirals}
sample. In this sample we combine all three possible spiral
classifications into a single classification. This is useful for
studies that require just a simple split into elliptical and
spiral samples. All the subsequent analysis is performed on both
the separated spirals (SS) and combined spirals (CS) samples.

We are then in a position to create the unweighted (UW) final sample. The simplest method involves giving
all classifiers equal weight and simply calculating the distribution of
classification for each galaxy. This distribution can now be
interpreted in a Bayesian manner: it represents our state of knowledge
about that particular galaxy.

\subsection{Weighted sampling}

The unweighted method discussed above does not discriminate between results from those who think carefully before classifying each galaxy, and those who take less time. Neither does it distinguish between the ability of our classifiers. It may therefore make sense to attempt to identify particularly `good' users. The meaning of `good' is naturally subjective, but one obvious strategy is to pay more attention to classifications from users who tend to agree with the majority. For this analysis, each user of the website was initially assigned a unit weight, as in the unweighted sample described above. A preliminary classifcation could then be obtained as before for each galaxy. The weighting assigned to individual users could then be adjusted according to how they agree with this assesment. A new set of galaxy classifications could then be prepared using the new weights, and the process repeated until the weights converge. 

In order to avoid the weightings assigned to users being distorted by the fainter end of our galaxy sample, we used only galaxies with petrosian radius $r_p>4.5$~arcsec and $r<$17 for our weighting. This leaves 257,000 galaxies involved in producing user weightings, although the resulting weightings are applied to all galaxy classifications. 

The algorithm used was as follows: let the weight of a user, $k$, be $w_k$ and set all initial weights to 1. We then integrate the database to find $h_i\left(j\right)$, the number of users who classified galaxy $i$ as being class $j$ (elliptical, anticlockwise spiral etc...). $N_g\left(k\right)$ is the number of galaxies classified by user $k$. The weights and $h_i$ are then updated by using the formul\ae : 

\begin{equation}
h_i\left(j\right)=A \sum_{\left(\mathrm{k = users\,who\,voted\,j\,for\,galaxy \,i}\right)}w_k,
\label{eq:2}
\end{equation}

where A is chosen so that the mean user weight is one, and 

\begin{equation}
   w_k = \sum_{i} \frac{h_i\left(\mathrm{j\,chosen\,by\,user\,k\,for\,galaxy\,i}\right)}{N_g\left(k\right)}.
\label{eq:1}
\end{equation}

This process can then be repeated until covergence. The final product is the weighted sample of galaxy classifications and a set of user weights.

It should be noted, however, that the process of reweighting favours the majority opinion. A user that is most similar to other users will get
upweighted and an user that does not conform to the pattern will get
downweighted. However, the overall agreement between users does not
necessarily mean improvement as people can agree on a wrong
classification. These effects must be calibrated using comparison with
standardized observations as described in Section \ref{sec:comp}.

\begin{figure}
\includegraphics[width=0.45\textwidth]{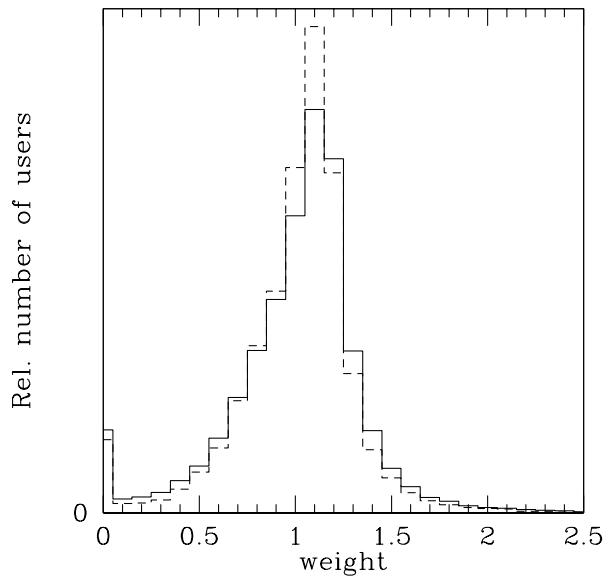}
\caption{The distribution of user weights for the separated (solid line) and combined (dashed line) data sets. The distribution for the separated spirals is slightly wider than that for the separated spirals sample.}\label{fig:weights}
\end{figure}

The distribution of user weights for both the separated and combined spiral samples is shown in Figure \ref{fig:weights}. Both distributions are slightly skewed toward the low-weighted end, and the combined spiral distribution is tighter than that for the seperated spiral data. This reflects the fact that as there are fewer possibilities to chose from in classifying a galaxy, for a set number of classifications better signal to noise is obtained, allowing us to better constrain the user weights.  

There are thus four possible combinations of separated spirals/Combined Spirals and Weighted/Unweighted samples. Unless otherwised stated, we use the weighted sample.  For
each sample we distill the data further into clean and superclean 
samples.
The galaxy is in a clean or superclean sample if  it has more than ten votes (in practice, this applies to almost our entire sample) and if 80\% or 95\% of users (or user weights in the case of the weighted sample) respectively agree on its type. These are extremely strong limits; an 80\% agreement is the equivalent of a 5-$\sigma$ detection for a galaxy with ten classifications, or a 10-$\sigma$ detection for a galaxy with the mean number of classifications. 

\begin{figure}
Class 1
\includegraphics[angle=90,width=0.4\textwidth]{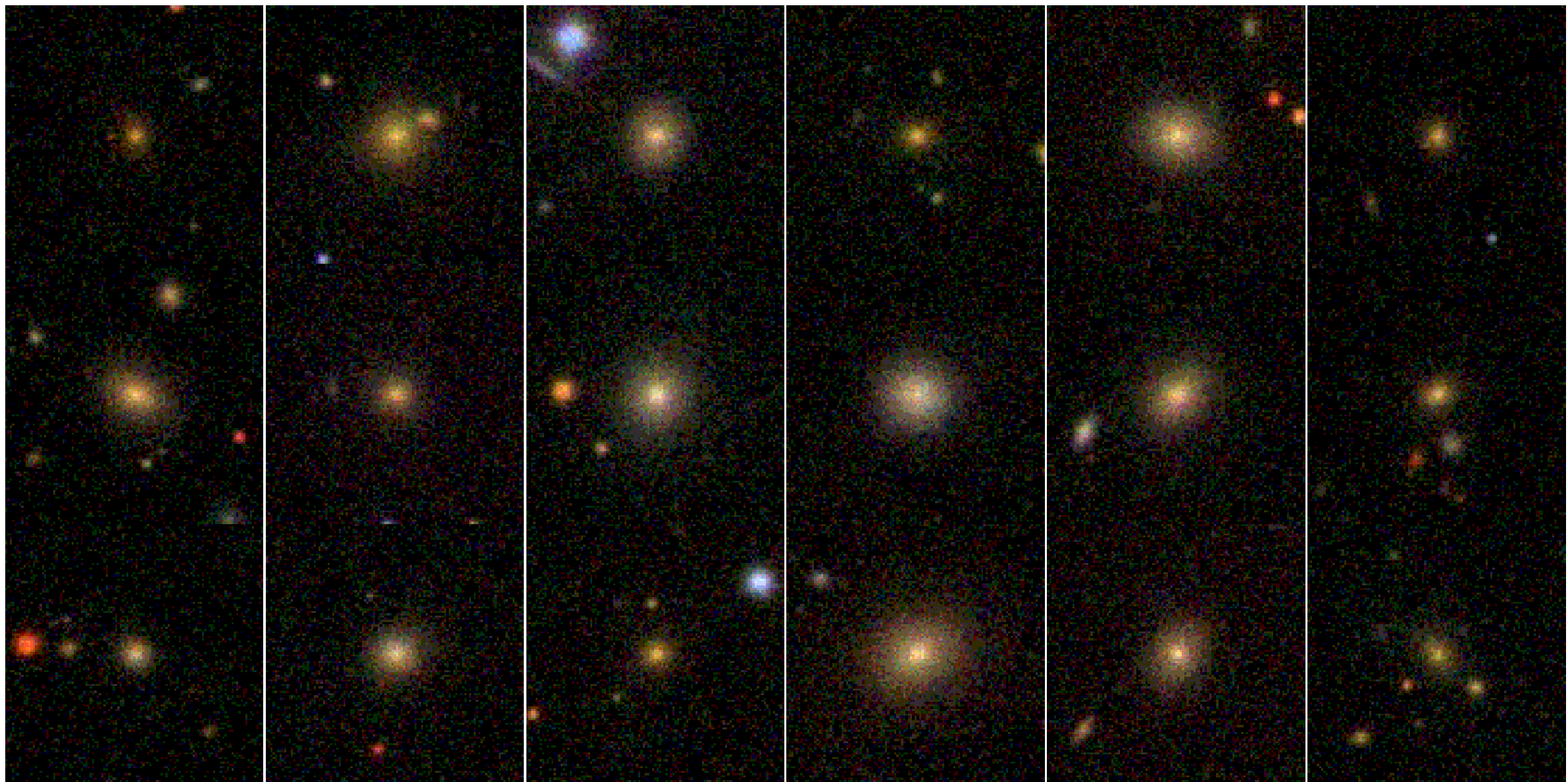}

\vspace*{0.3cm}

Class 2
\includegraphics[angle=90,width=0.4\textwidth]{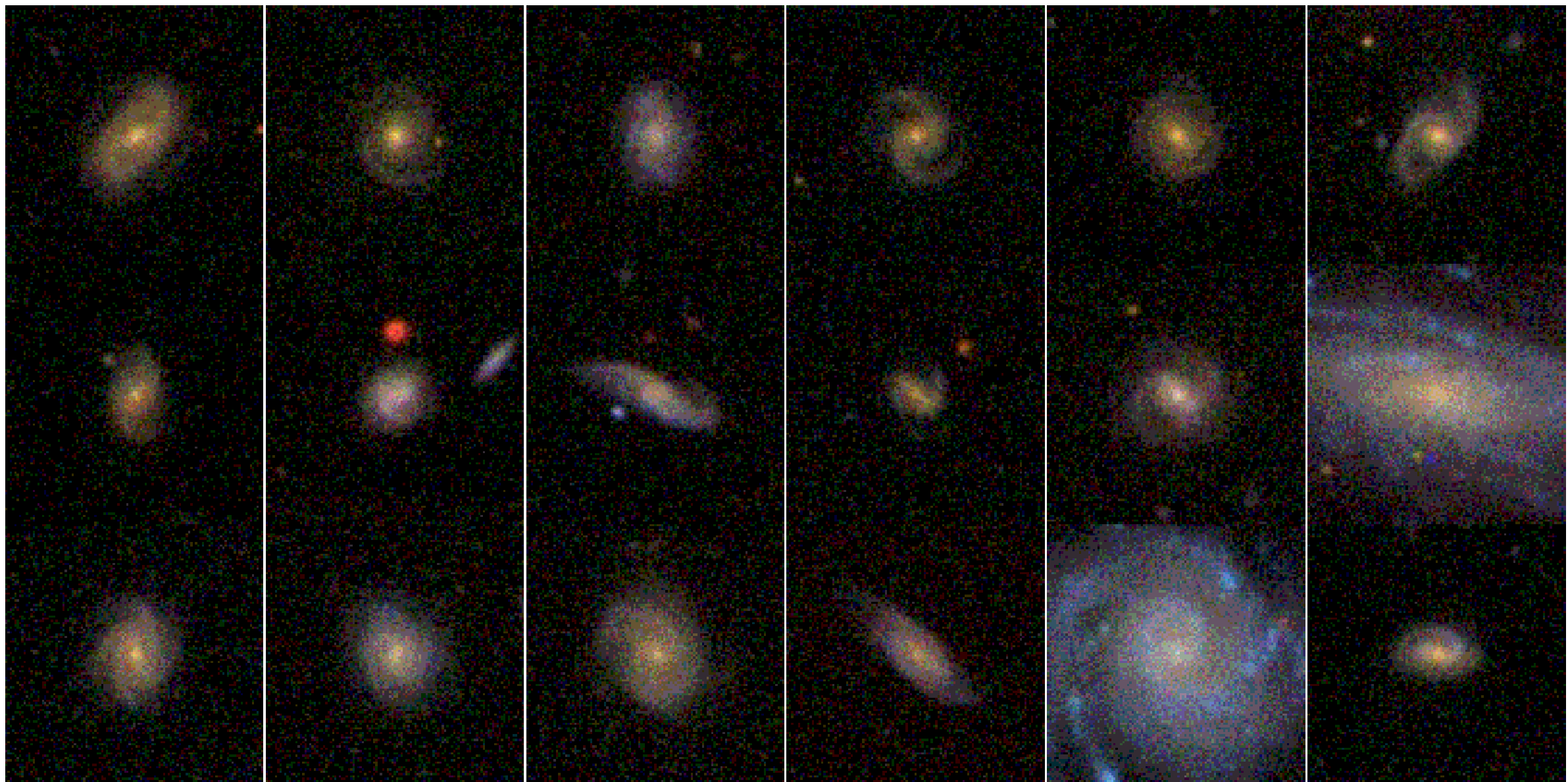}

\vspace*{0.3cm}

Class 3
\includegraphics[angle=90,width=0.4\textwidth]{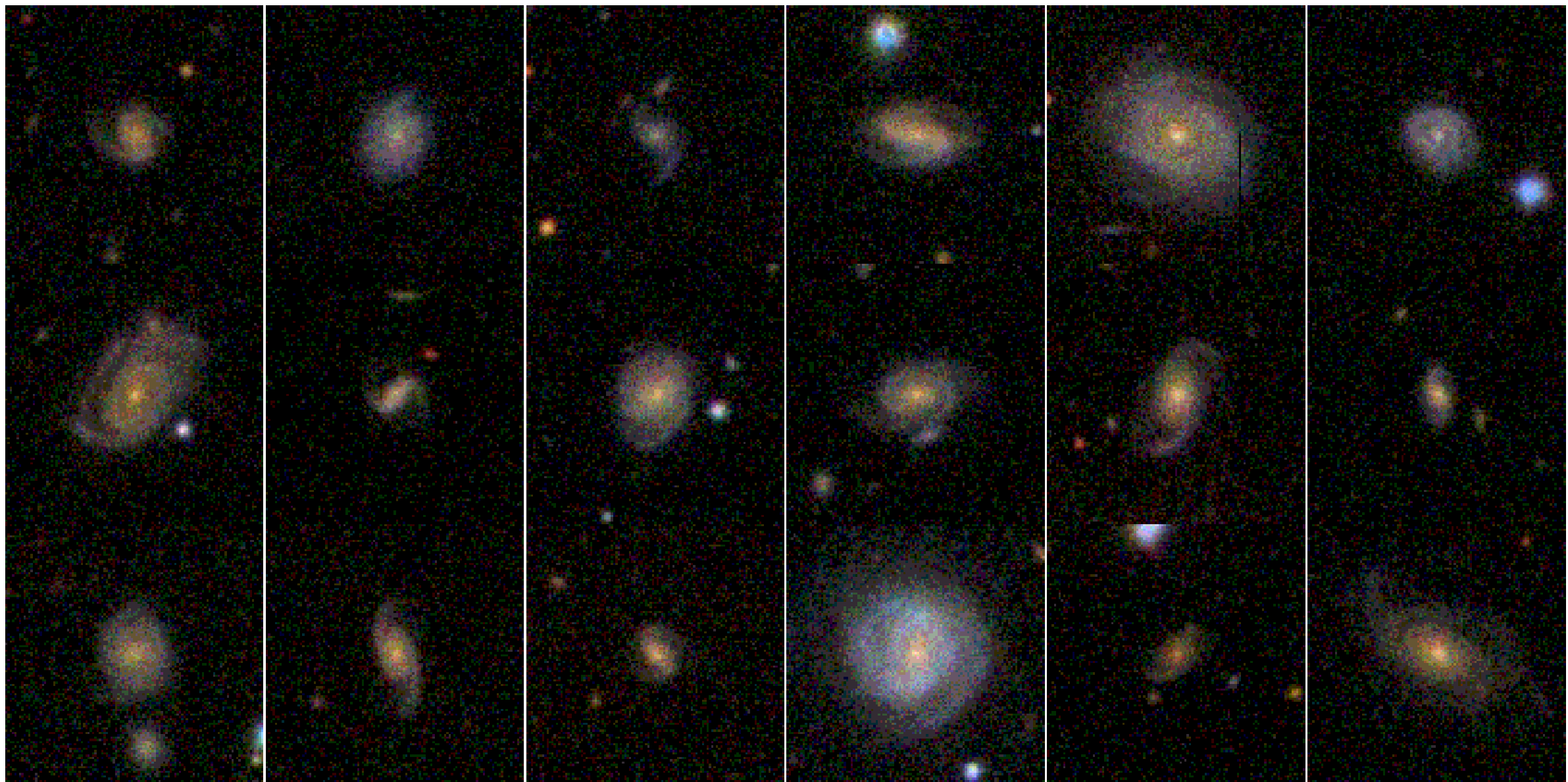}

\vspace*{0.3cm}

Class 4
\includegraphics[angle=90,width=0.4\textwidth]{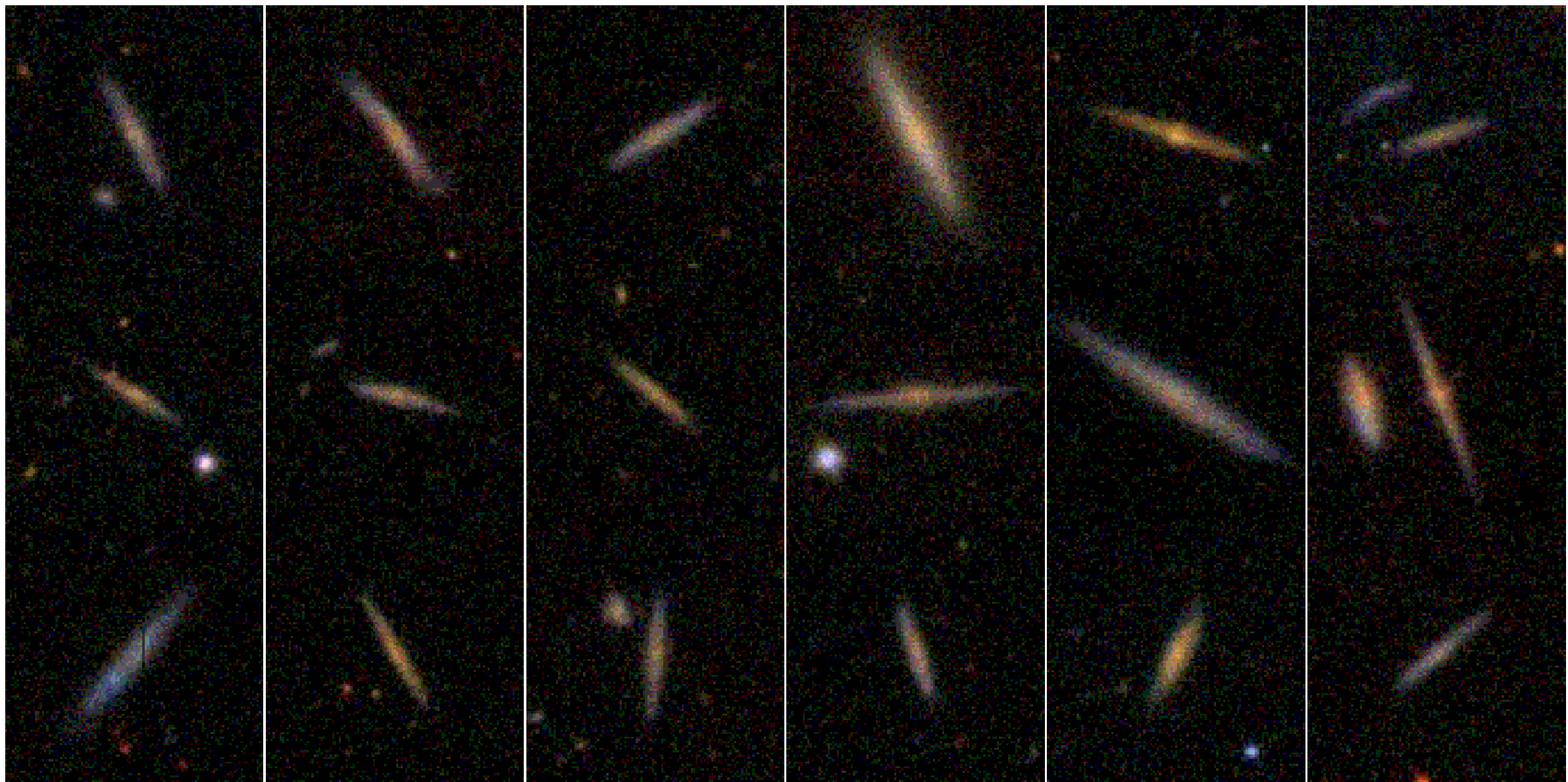}

\vspace*{0.3cm}

Class 5
\includegraphics[angle=90,width=0.4\textwidth]{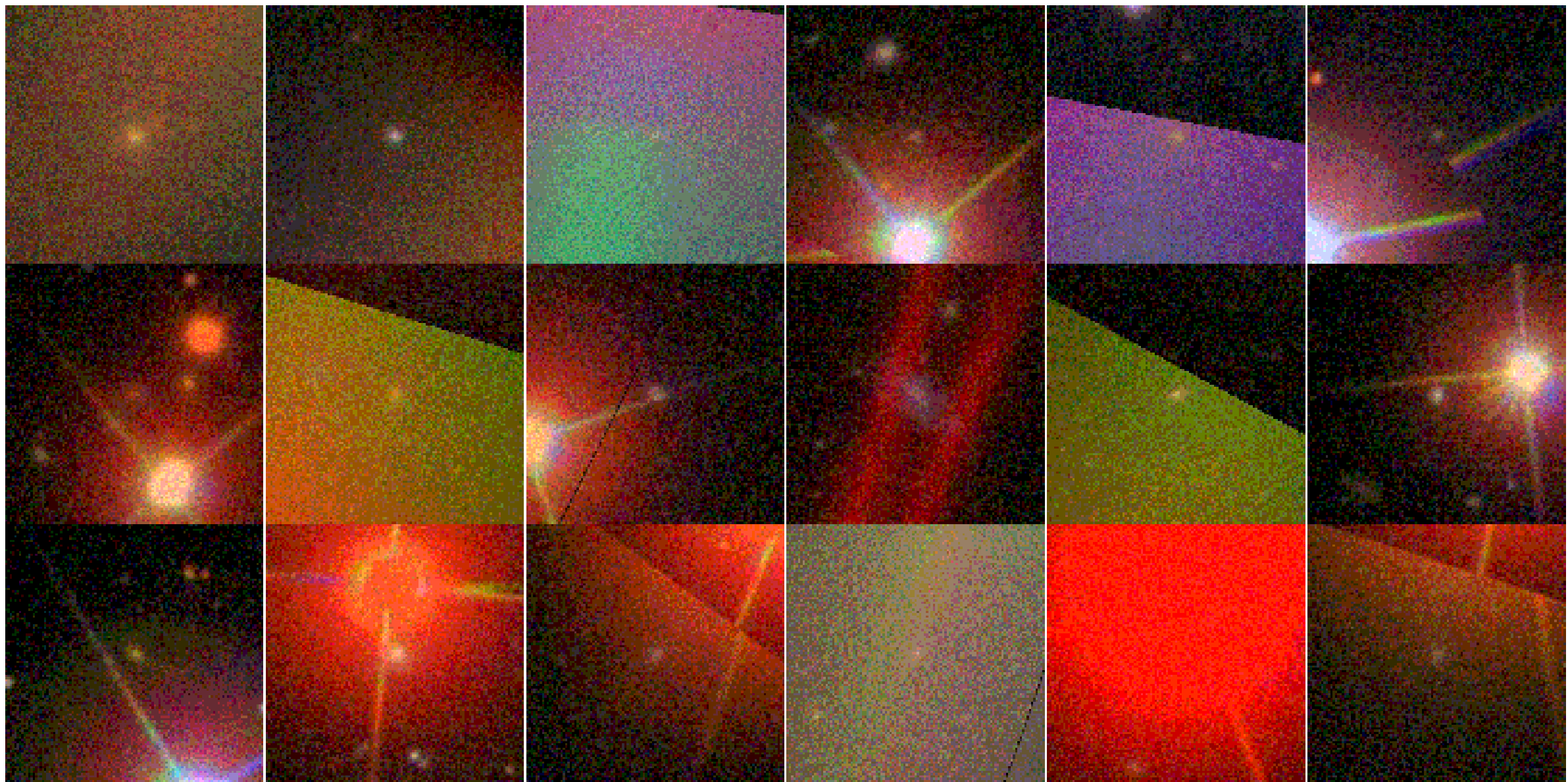}

\vspace*{0.3cm}

Class 6
\includegraphics[angle=90,width=0.4\textwidth]{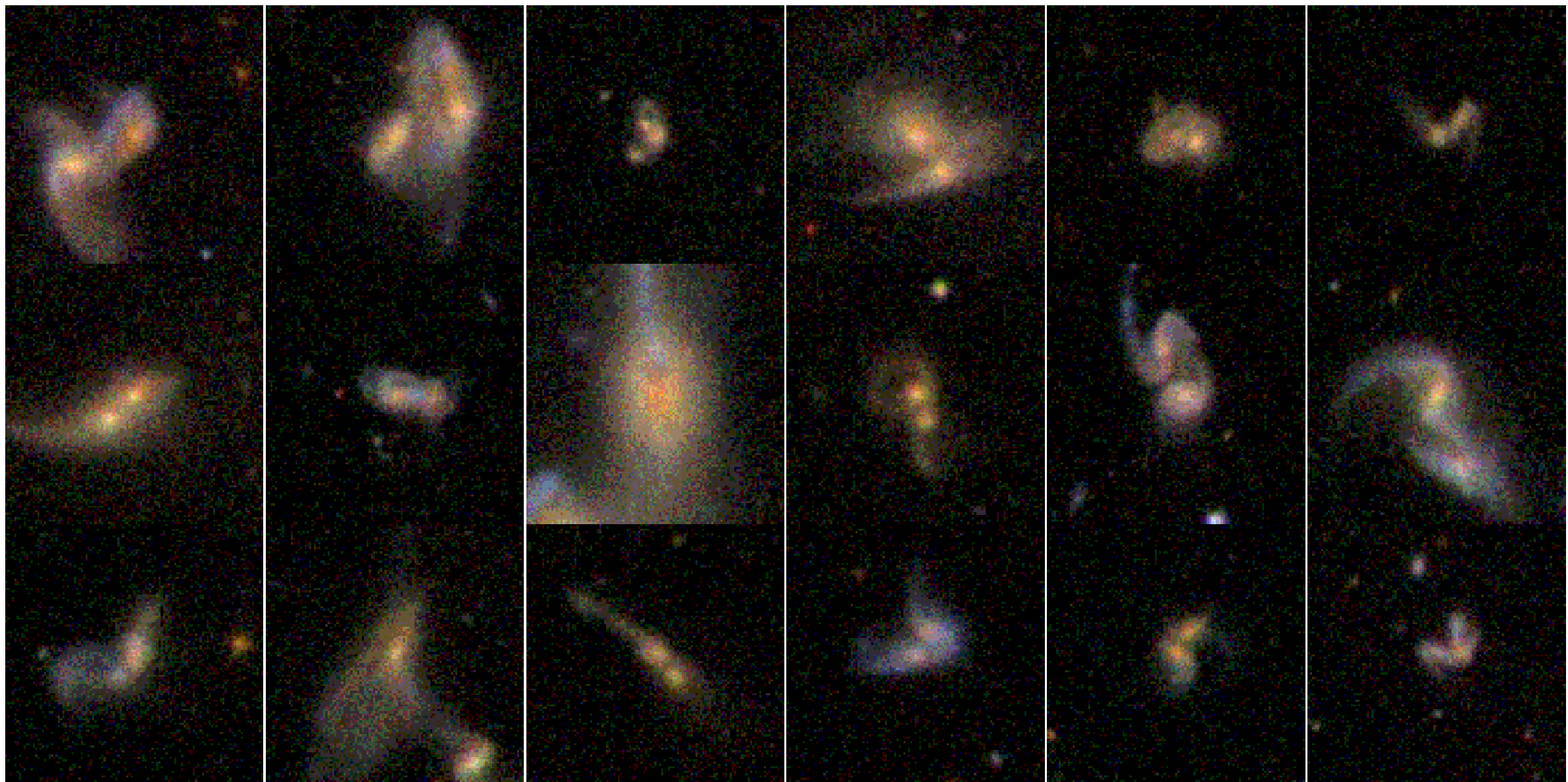}
\caption{Examples of galaxies in each class drawn from the weighted superclean sample. Each image is 51.2 x 51.2 arcsec.}\label{fig:scegs}
\end{figure}

\begin{figure}
Class 1
\includegraphics[angle=90,width=0.4\textwidth]{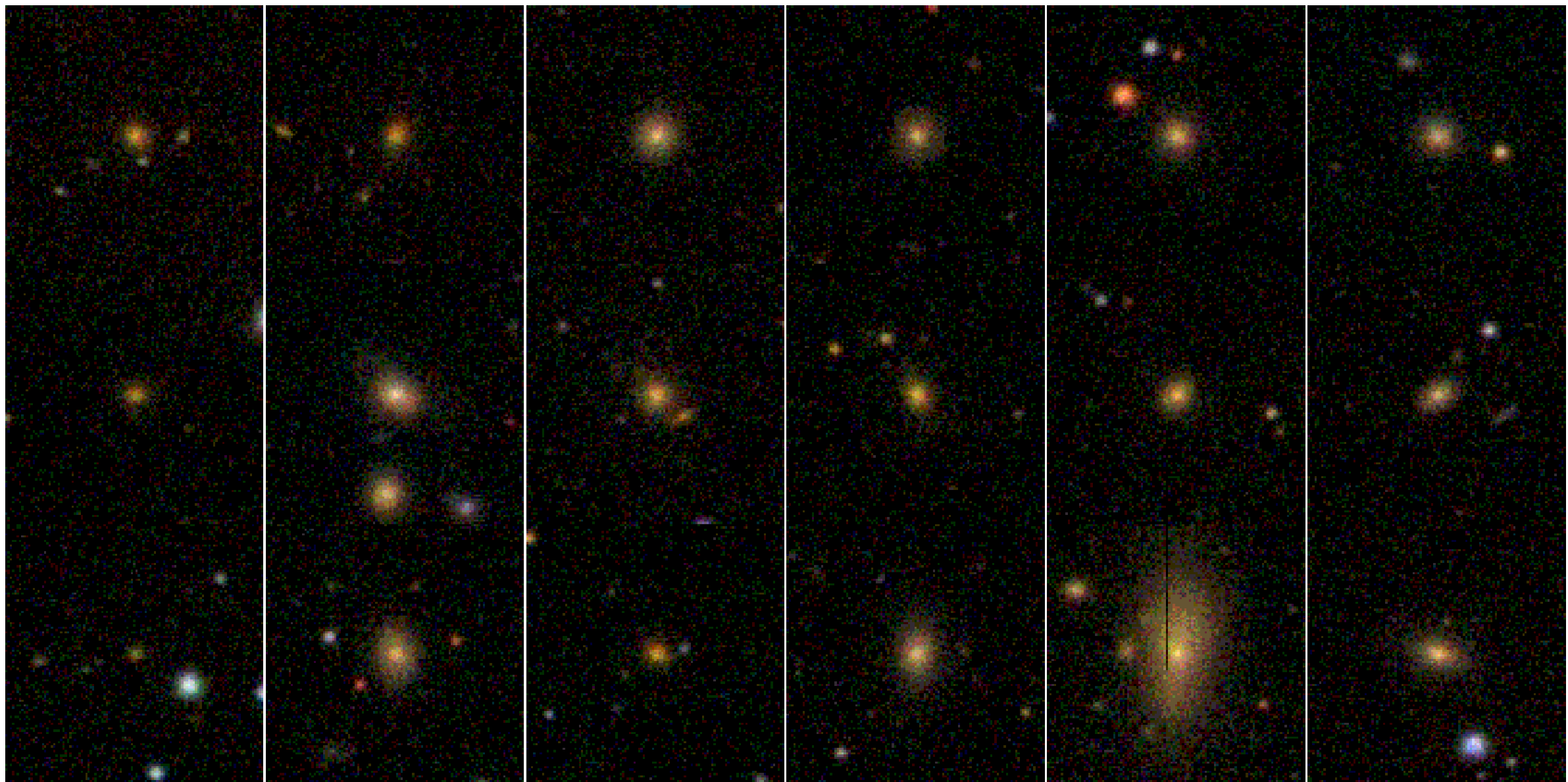}

\vspace*{0.3cm}

Class 2
\includegraphics[angle=90,width=0.4\textwidth]{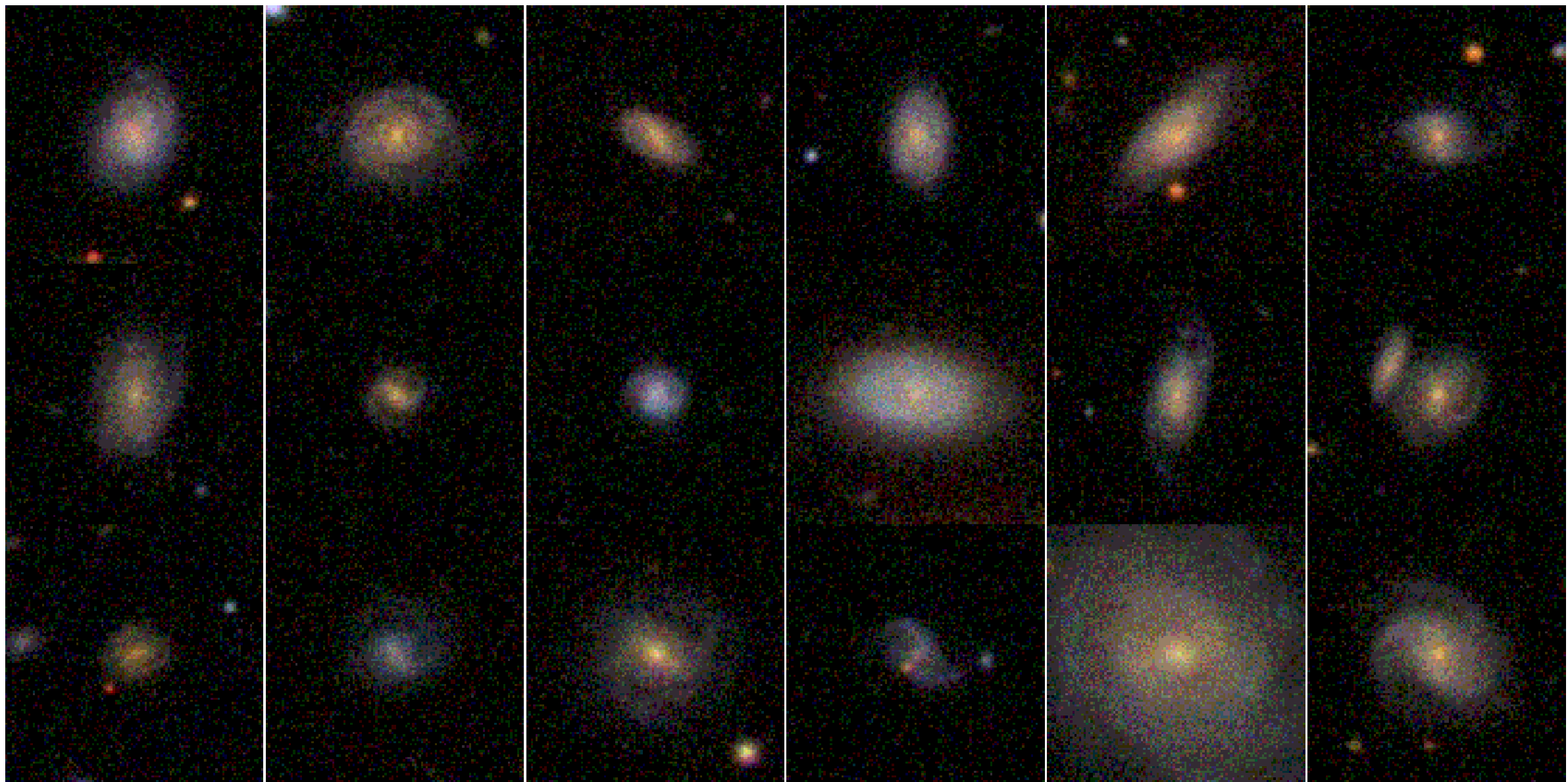}

\vspace*{0.3cm}

Class 3
\includegraphics[angle=90,width=0.4\textwidth]{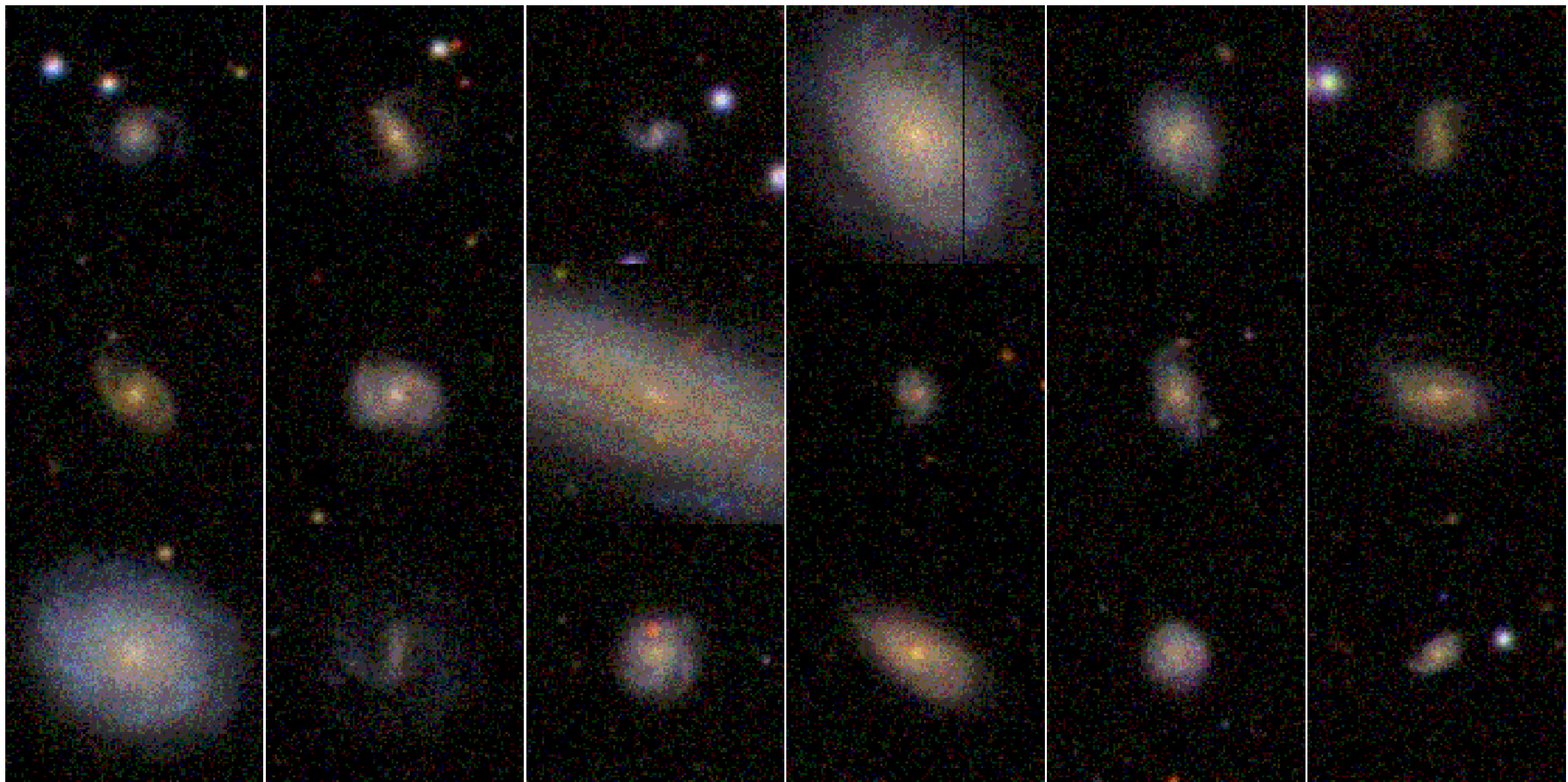}

\vspace*{0.3cm}

Class 4
\includegraphics[angle=90,width=0.4\textwidth]{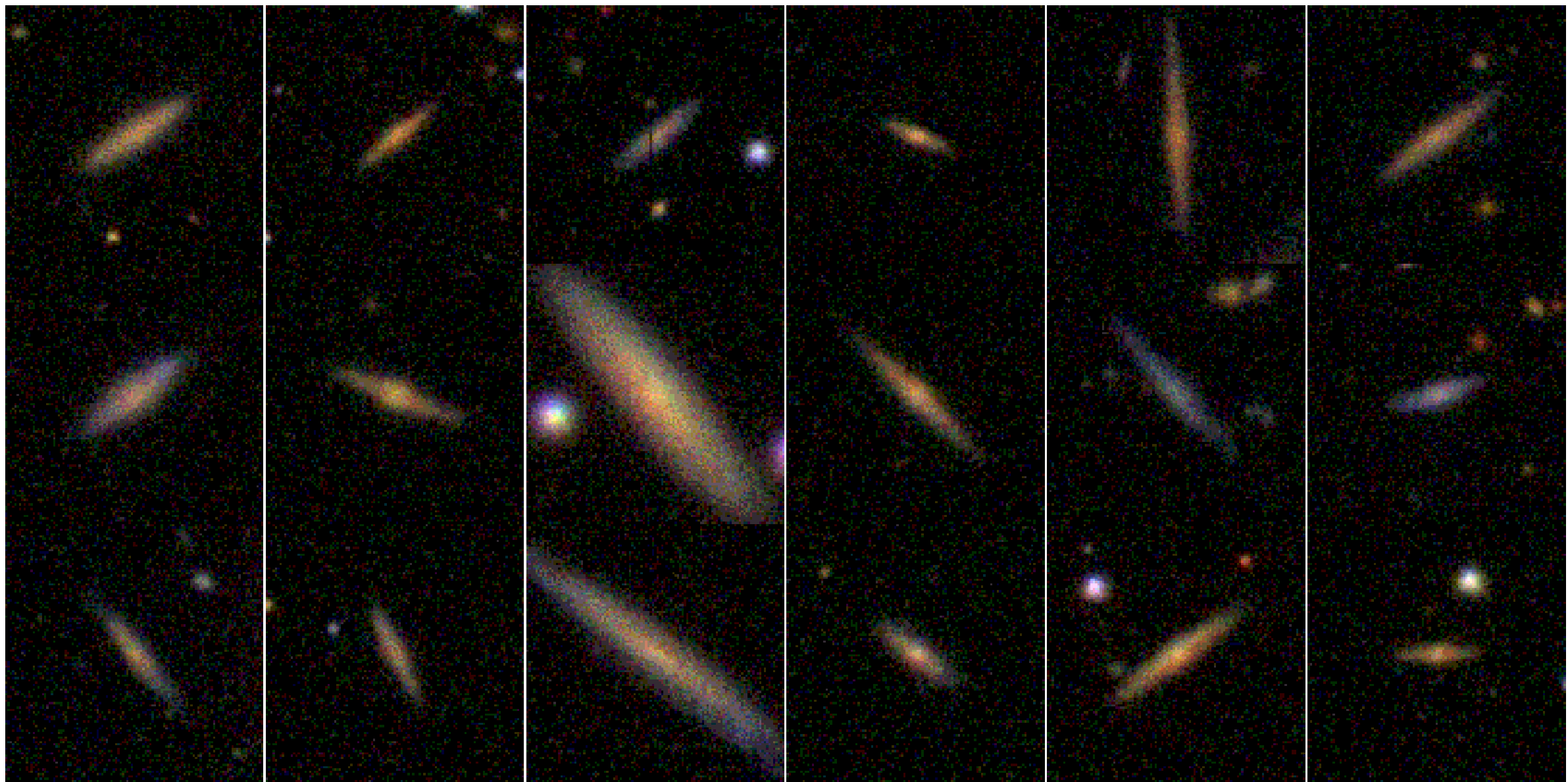}

\vspace*{0.3cm}

Class 5
\includegraphics[angle=90,width=0.4\textwidth]{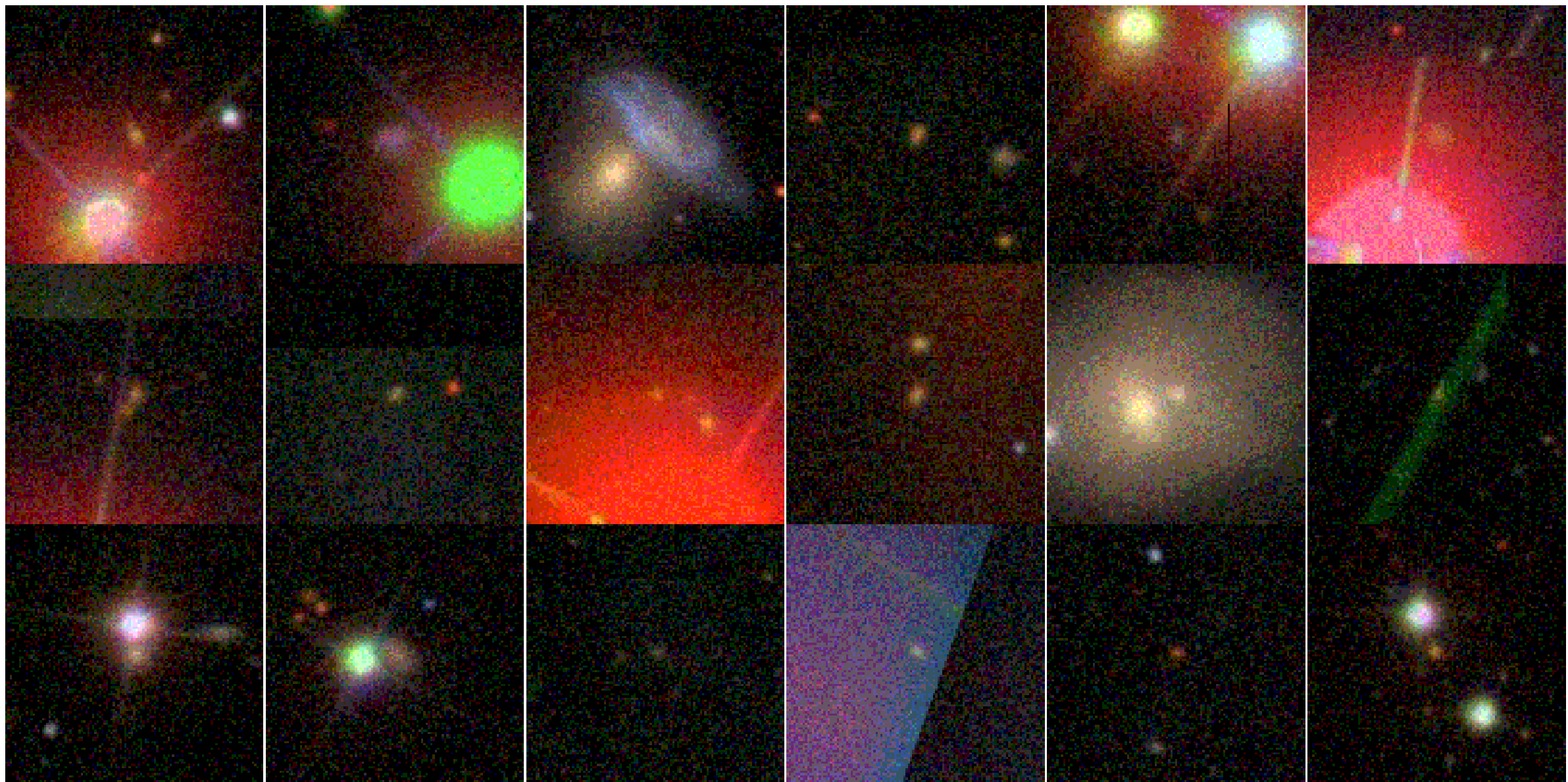}

\vspace*{0.3cm}

Class 6
\includegraphics[angle=90,width=0.4\textwidth]{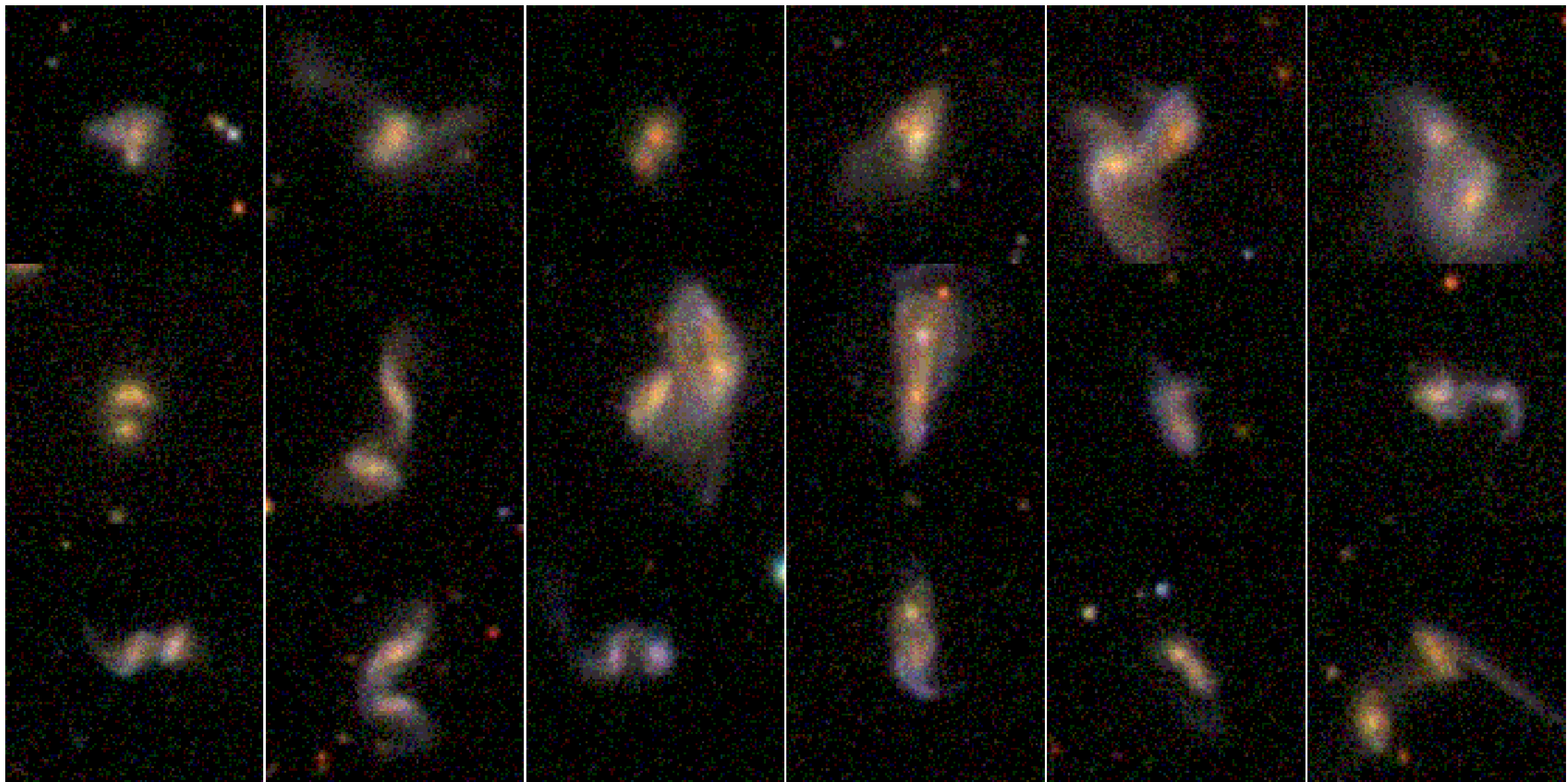}
\caption{Examples of galaxies in each class drawn from the weighted clean sample. Each image is 51.2 x 51.2 arcsec.}\label{fig:cegs}
\end{figure}

Examples of objects in each category randomly extracted from the weighted superclean sample are shown in Figure \ref{fig:scegs} and examples from the clean sample in Figure \ref{fig:cegs}. It should be noted that the combined spiral sets cannot be recovered simply by taking the single spirals clean set and combining classes 2, 3 and 4. For example, a galaxy that has all its votes evenly split between classes 2 and 3 (clockwise and anticlockwise spirals) will definitely be included in the combined spiral clean set, but would not appear in the separated spirals clean set.

The effect of this weighting process is shown in Table \ref{tab:weight1} for the separate spirals and in Table \ref{tab:weight2} for the combined spirals. These tables show that in all cases the vast majority (above 99\% in all but two cases) of classifications made in the unweighted samples are carried forward into the weighted sample. This means that the weighting described above is not changing classifications. However, extra galaxies are included in each classification in the weighted sample. This effect is largest ($\sim 15$\%) for the elliptical classes; this can easily be explained by the fact that it is more difficult to agree on the presence of spiral structure than on an elliptical morphology. A smaller proportion of spiral systems reach the stringent criteria for inclusion in the superclean sample. In total, more than 300,000 galaxies are included in the most inclusive sample; this is the largest sample of morphologically clasified galaxies by a factor of 10. 

\begin{table*}
\begin{tabular}{cccccc}
sample & class & \# in weighted & \# in unweighted & \% increase & \%
 common \\
\hline
\hline
clean & elliptical & 219326 & 184743 &  15.8 &  99.98 \\
\hline
clean & cw spiral & 17571 & 17100 &   2.7 &  99.70 \\
\hline
clean & acw spiral & 18946 & 18471 &   2.5 &  99.72 \\
\hline
clean & other spiral & 27310 & 26037 &   4.7 &  99.46 \\
\hline
clean & star/don't know & 8134 & 8074 &   0.7  &  99.75 \\
\hline
clean & merger & 1062 & 961 &  9.5 &  99.48 \\
\hline
\hline
superclean & elliptical & 26200 & 19121 &  37.0 &  99.7 \\
\hline
superclean & cw spiral & 6532 & 6106 &  7.0 &  99.4 \\
\hline
superclean & acw spiral & 7486 & 7034 &  6.4 &  99.4 \\
\hline
superclean & other spiral & 4760 & 4247 &  12.1 &  99.2 \\
\hline
superclean & star/don't know & 5589 & 5393 &  3.6 &  99.5 \\
\hline
superclean & merger & 70 & 62 &   12.9 &  96.8 \\
\hline
\hline
\end{tabular}
\caption{Comparison of classification between weighted and unweighted samples. For each class the Table shows the number of galaxies so classified, the percentage increase in weighted over unweighted classifications and the percentage of the unweighted sample in common with the weighted sample.}\label{tab:weight1}
\end{table*}

%\vspace*{3cm}
\begin{table*}
\begin{tabular}{cccccc}
Combined Spirals\\
sample & class & \# in weighted & \# in unweighted & \% increase & \%
 common \\
\hline
\hline

clean & elliptical & 208437 & 184743 &  12.8 &  99.9 \\
\hline
clean & spiral & 101855 & 97848 &   4.1 &  99.9 \\
\hline
clean & star/don't know & 8126 & 8074 &   0.6 &  99.9 \\
\hline
clean & merger & 1056 & 961 &  9.9 &  99.5 \\
\hline
\hline
superclean & elliptical & 23806 & 19121 &  24.5 &  99.7 \\
\hline
superclean & spiral & 34673 & 32559 &  6.5 &  99.7 \\
\hline
superclean & star/don't know & 5573 & 5393 &  3.3 &  99.7 \\
\hline
superclean & merger & 67 & 62 & 8.1 &  96.8 \\
\hline
\hline
\end{tabular}
\caption{As table \ref{tab:weight1}, but for the combined spirals data set.}\label{tab:weight2}
\end{table*}

Inspection of Tables \ref{tab:weight1} and \ref{tab:weight2} immediately reveal that many more galaxies in the clean sample have been classified as elliptical than spiral. The elliptical:spiral ratio is $\sim 3$ for both the weighted and unweighted clean sample. The combined spiral clean sample produces a much lower ratio ($\sim 2$). This difference is another illustration of the discrimination against spirals discussed in the previous paragraph. The combined spirals data should be free of such effects, but still has a large elliptical fraction. This reflects the tendency of our users to classify objects which are faint, small in angular extent on the sky or both as elliptical if no spiral features are present. It is therefore important to apply magnitude cuts to the data before using data for the population as a whole; individual users of the Galaxy Zoo data will require different cuts and so we do not impose any on the clean sample ourselves. The `true' elliptical fraction for a volume-limited sample of well-classified galaxies is discussed in Section \ref{sec:efrac}. 

\section{Comparision with other samples}\label{sec:comp}

In order to assess the reliability of the Galaxy Zoo classifications, we compare our sample with that produced by previous projects. The MOSES sample (Schawinski et al., 2007) described in Section 1 consists of 15729 galaxies classified as elliptical selected from an initial set of 48023 galaxies. Of the 48023 the clean sample includes 
classifications for 19649 systems.  The results for the weighted clean sample are given in Table \ref{tab:MOSES}

%\centerline{\begin{minipage}{8cm}
\begin{table}
\begin{tabular}{|c|cc|c|}
\hline
 & moses e & moses other & moses all \\
\hline
Elliptical & 10,858  & 1,676 & 12,534 \\
ACW spiral  &  0 & 2,493 & 2,493\\
CW spiral &  2 & 2,598 & 2,600  \\
Other spiral  &  4 &  1,940 & 1,944\\
Star/don't know  & 0 & 4& 4 \\
Merger &  4 & 70& 74  \\
\hline
tot  & 10,868 & 8,781  & 19,649 \\
\hline
\end{tabular}
\label{tab:MOSES}
\caption{Comparision of classifications for galaxies in both MOSES and the Galaxy Zoo weighted clean sample. Most MOSES ellipticals are classified by Galaxy Zoo as elliptical.}
\end{table}
%\end{minipage}

More than 99.9\% of the galaxies classified as MOSES ellipticals which are in the Galaxy Zoo clean sample are found to be ellipticals by Galaxy Zoo. However, $\sim 15$\% of the ellipticals included in both the Galaxy Zoo clean sample and MOSES were not classified as elliptical by MOSES. All MOSES ellipticals in the superclean sample are classified by Galaxy Zoo as ellipticals, but the sample contains 3\% more ellipticals than MOSES. These extra ellipticals are the result of the different motivation of the studies; MOSES was an attempt to produce a very clean set of ellipticals, whereas the Galaxy Zoo samples include more of the S0-Sa continuum in the resulting sample. The Galaxy Zoo instructions to volunteers did not mention disks at all, and so galaxies which are elliptical in morphology but have visible disks would have been included in Galaxy Zoo but not in MOSES. 

\begin{figure}
\includegraphics[width=0.45\textwidth]{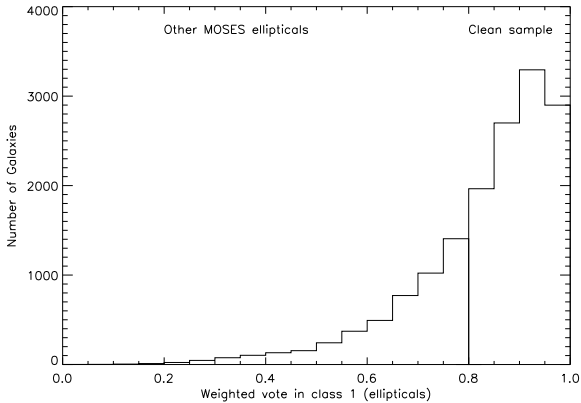}
\caption{Weighted vote in class 1, corresponding to ellipticals, for the 15729 galaxies classified in the MOSES sample as elliptical. Those with a weight in this class greater than 80\% are included in the Galaxy Zoo clean sample, but it is clear from this figure that this is an effectively arbitary cut-off point. For approximately 90\% of the galaxies, the majority of weighted votes are in the elliptical category.}\label{fig:mosese}
\end{figure}

There are also ellipticals which are in the MOSES catalogue but not in the clean sample.  The distribution of weighted votes for MOSES ellipticals including both those included in the clean sample and those which are not is shown in Figure \ref{fig:mosese}. The majority of weighted votes in almost all cases support an elliptical classification. The requirement for the clean classification of a weighted vote of 80\% thus lies in the middle of a continuous distribution of weights. In most cases, the remaining votes show that a small minority of users selected other options, usually for good reasons such as the presence of a nearby satellite trail or some evidence of a disturbed morphology. The weighted vote in the spiral categories is below 20\% in all but an insignificant number of cases. This example thus illustrates the stringency of the clean and superclean samples; only galaxies on which a large majority of users agree are included in the final samples. 

\begin{figure}
\includegraphics[angle=270,width=0.45\textwidth]{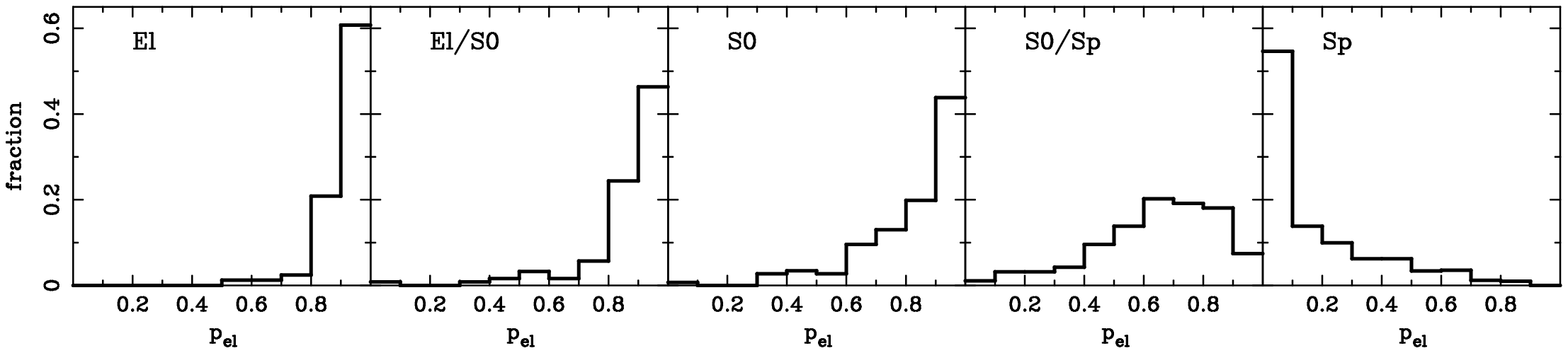}
\includegraphics[angle=270,width=0.45\textwidth]{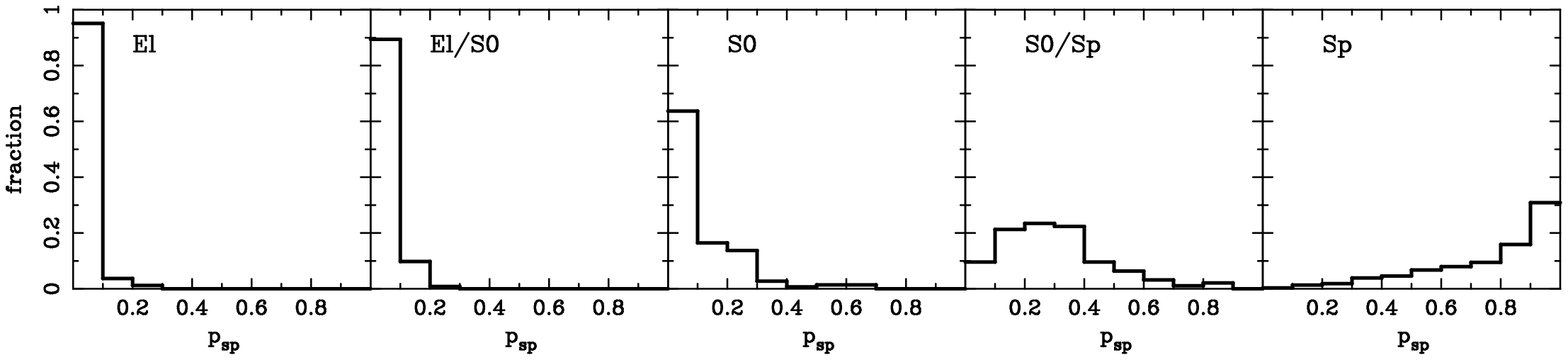}
\caption{Histographs showing a comparision between Galaxy Zoo classifications and those from Fukugita et al. The axis plots the fraction of the weighted vote in the clean sample that Galaxy Zoo allocated to elliptical (top) and spiral (bottom) for those galaxies classified by Fukugita et al. as elliptical (el) as ellipitcal/s0 (El/S0), S0, S0/Spiral (S0/Sp) and Spiral (Sp).}\label{fig:fuk}
\end{figure}

In order to investigate this effect and provide an independent check on the data, we consider another set of SDSS galaxies, those classified by Fukugita et al. (2007). They use the statistic `T' as their classification, taken from an average of 3
classifiers (rounding to the nearest half integer). The options available are:\\0 (E), 1 (S0), 2(Sa), 3(Sb), 4(Sc), 5(Sd), and 6(Im). Unlike MOSES, therefore, we can use this smaller sample to probe the response of Galaxy Zoo users to galactic morphologies of a wide variety of sub-types. 
$1\le T \le  5$ are `spiral' systems, $T=0,0.5$ are
`elliptical' and $T<0$ and $T=6$ are unclassified or irregular systems. 
Of their sample of 2275 galaxies we have clean classifications for 
1300 galaxies (621 are included in the superclean sample). The mean T for galaxies included in the clean sample and classified as elliptical is 0.52, and that for spirals 2.54. A full comparision for the clean sample is given in Table \ref{tab:Fuk}, and the distribution of weights shown in Figure \ref{fig:fuk}.

%\vspace{1.cm}
%
%\centerline{\begin{minipage}{14cm}
%\centering
\begin{table*}
\begin{tabular}{|c|cccccccccccccc|c|}
\hline

T=&	$< 0$&	0&	0.5&	1&	1.5&	2&	2.5&	3&	3.5&	4&	4.5&	5&	5.5&	6& all\\
Elliptical&	1&	267&	190&	170&	41&	11&	2&	0&	0&	0&	0&	0&	0&	0&	682\\
Spiral&	3&	0&	0&	0&	5&	21&	71&	136&	151&	160&	38&	13&	5&	2&	605\\
Star/don't know&	0&	1&	0&	0&	0&	0&	0&	0&	0&	0&	0&	0&	0&	0&	1\\
Merger&	5&	0&	1&	0&	0&	1&	1&	2&	0&	1&	1&	0&	0&	0&	12\\
\hline
Total &  9&	268&	191&	170&	46&	33&	74&	138&	151&	161&	39&	13&	5&	2&   1300\\

\hline
\end{tabular}

\begin{tabular}{|c|cccccccccccccc|c|}
\hline

T=&	$< 0$&	0&	0.5&	1&	1.5&	2&	2.5&	3&	3.5&	4&	4.5&	5&	5.5&	6& all\\	
\hline
Elliptical&	0&	145&	88&	52&	3&	0&	0&	0&	0&	0&	0&	0&	0&	0&	288\\
Spiral&	0&	0&	0&	0&	1&	7&	22&	58&	94&	119&	24&	6&	2&	0&	333\\
Star/don't know&	0&	0&	0&	0&	0&	0&	0&	0&	0&	0&	0&	0&	0&	0&	0\\
Merger&	0&	0&	0&	0&	0&	0&	0&	0&	0&	0&	0&	0&	0&	0&	0\\
\hline
Total&	0&	145&	88&	52&	4&	7&	22&	58&	94&	119&	24&	6&	2&	0&	621\\
\hline
\end{tabular}
\caption{Comparision of the combined spirals clean (top) and superclean (bottom) sample results with those from Fukugita et al. 2007. Their classification is given on the x-axis, and the Galaxy Zoo results on the y-axis. See table \ref{tab:buttons} for details of our classification system.}\label{tab:Fuk}
\end{table*}
%\end{minipage}

 The vast majority of galaxies classified as elliptical in the clean sample are classified as elliptical (T=0,0.5) by Fukugita et al. Of the ellipticals in the clean sample 92\% correspond to early-type galaxies in the Fukugita et al. sample (S0 or ellipticals). The equivilent figure for the superclean sample is 99\%. All but two of the remaining galaxies classified by Galaxy Zoo as ellipticals were classified by Fukugita et al. as Sa. This supports the hypothesis that the excess of ellipticals seen when comparing to the MOSES sample is composed mostly of Sa galaxies; astronomers are more reluctant than the general public to classify something with a definite disk as an elliptical galaxy. No mention of disks was made in the instructions to our classifiers, but such an addition is an obvious change to make in future versions of the website. For the ellipticals, we find no obvious trend between T and magnitude, but T appears to be correlated with the weight of the classification. Switching to the superclean sample therefore improves the correlation between the samples. 

Finally, \citet{Longo} selected the spiral galaxies used in his study by visual inspection of galaxies in the SDSS. Of the 2834 galaxies in this sample, 2498 are included in the Galaxy Zoo clean sample, 2491 of which are classifed as spirals. The other seven are classified as mergers (4) or as `star/don't know' (3). A comparision with the clean separate spiral catalogue finds excellent agreement for the winding sense of the spiral arms (99.6\%). In the 10 cases where there was a disagreement, further inspection reveals that the disagreement can in each case be put down to human error in the catalogue of Longo (2007), illustrating the advantage of obtaining multiple independent classifications for each system.

The three data sets with which we have compared Galaxy Zoo were compiled in very different ways to test different hypotheses. However, in each case we find a remarkable degree of agreement (better than 90\% in most cases) between our data set and those compiled by professional astronomers. We can therefore conclude that using data from volunteers will not substantially degrade the quality of the resulting catalogue while expanding the number of classified galaxies by a large factor. 

\subsection{Measuring bias}
\label{sec:biasresults}

The aim of the Galaxy Zoo study was to produce a catalogue of morphologically selected galaxies, independent of the bias introduced by using proxies for morphology. Potentially the strongest of these biases is the correlation between morphology and colour. While the instructions to users did not include any mention of colour, it is a fact that most spirals are significantly bluer than most elliptical galaxies, and this fact will be quickly learnt by classifiers. It is therefore possible that our selection will include a residual colour bias. In order to quantify the size of any such effect, a programme of bias testing was undertaken. Users were shown either a mirror image of the original data, or a monochrome image produced from the coloured images. (These images are not single filter images, but rather a black and white version of the colour image produced by the SDSS pipeline as described above). The results are given in Table \ref{tab:bias}. 

Any bias study such as this runs the risk of changing the behaviour of those taking part itself, a phenomenon known in social science as the Hawthorne effect \citep{Mayo,Adair}. To give just one example of how this might affect Galaxy Zoo, users may be more cautious with their classifications if they think that they are being tested for bias rather than just being asked to make their best guess.

%\begin{table}
%\begin{tabular}{|c|c|c|c|c|c|c|c|}
%\hline
%Original & & & Monochrome & & & Mirrored \\
%class & $<$ \% $>$ & $\sigma$  & $<$ \% $>$ & $\sigma$  & $<$ \% $>$ & $\sigma$\\
%\hline
%1 & 53.82546 & 0.12218 & 55.96226 & 0.12278 & 55.01500 & 0.12386 \\
%2 \& 3 \& 4 & 32.37073 &  0.13069 & 28.97221 &  0.12775 &  30.05011 &  0.13050\\
%5 & 10.11606 &  0.05747 & 11.05859 &  0.06038 &  11.26452 &  0.05991\\
%6 &  3.68775 &  0.04500 &  4.00695 &  0.05016 &  3.67036 &  0.04543\\
%\end{tabular}
%\caption{Results of the bias study}
%\end{table} 

\begin{table}
\begin{tabular}{|c|c|c|c|c|}
\hline
& Original &  Monochrome & Mirrored \\
class & $<$ \% $>$ ($\sigma$)  & $<$ \% $>$ ($\sigma$)  & $<$ \% $>$ ($\sigma$)\\
\hline
1 & 53.82 (0.12) & 55.96 (0.12) & 55.02 (0.12) \\
2 \& 3 \& 4 & 32.37 (0.13) & 28.97 (0.13) &  30.05 (0.13)\\
5 & 10.12 (0.06) & 11.06 (0.06) &  11.26 (0.06)\\
6 &  3.69 (0.05) &  4.01 (0.05) &  3.67 (0.05)\\
\end{tabular}
\caption{Results of the bias study. The numbers given are the average percentage of votes that each class receives per galaxy, with 1 sigma errors obtained from jackknife resampling (see Land et al. for details).}\label{tab:bias}
\end{table} 

A change in user behaviour between the original classifications and those collected as part of this bias study is indeed seen. In particularly, users are more careful in their classifications during the bias study. This effect makes it impossible to make a fair comparision between classifications made before the bias study started and those collected during it. However, we do not expect mirroring the images to influence the choice between spiral and elliptical galaxies, and we can thus use the mirrored images as a control. The result of a comparision between classifications of monochrome and mirrored images is a significant (of order 5-$\sigma$) difference in behaviour. Users shown monochrome images are more likely to classify a galaxy as an elliptical, and correspondingly less likely to classify a galaxy as a spiral. There is also a bias in favour of classifying a galaxy as a merger; this is presumably due to the loss of colour information which enables us to distinguish two seperate galaxies from one merging system. However, although these are statistically significant differences, they are small. The mean percentage of votes for the elliptical class increases from 55\% to 56\%, for example. We are thus justified in ignoring this bias when using the catalogues for most purposes.\label{sec:colors}

By using the monochrome images as a control, we can test for a bias in the classification of the direction of spiral arms. A significant bias in favour of anticlockwise classifications was found, and is discussed in Land et al. (2008). We also expect a bias toward elliptical galaxies for more distant systems as it becomes harder to resolve features which would indicate a spiral system. Providing a conservative cut in magnitude, size or redshift (or some combination of the three) is made, then this bias can be safely ignored. When considering the properties of the population as a whole, it is possible to be more quantitative in accounting for the effect of this bias on the results; for a full discussion of this technique, see Bamford et al. 2008.

\section{Colour-magnitude diagrams} 

\begin{figure}
\includegraphics[width=0.45\textwidth]{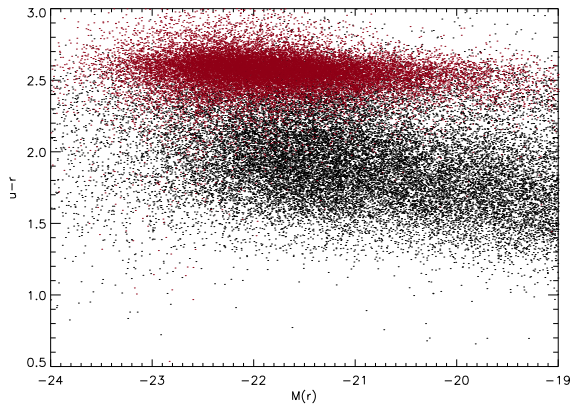}
\caption{Colour-magnitude diagram for galaxies in the weighted superclean combined spirals sample. Systems classified as spiral are shown in black, those classified as elliptical in red.}\label{fig:cmag}
\end{figure}

\begin{figure}
\includegraphics[width=0.45\textwidth]{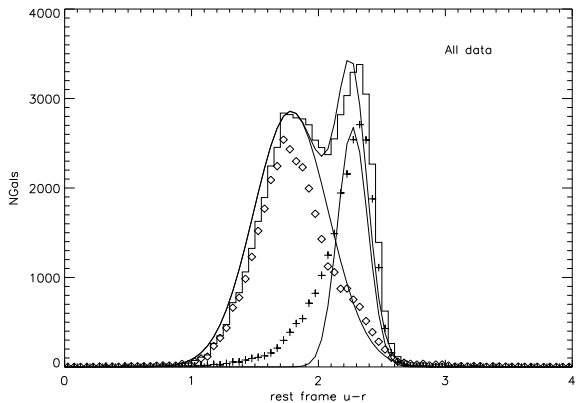}
\caption{Colour-magnitude histograms for galaxies in the clean combined spirals sample. Crosses mark ellipticals, diamonds spirals. A two-Gaussian fit to the complete data is shown (top), together with the individual Gaussians used in the fit. The curve shown is the limit for the main galaxy catalogue; objects below this line were drawn from the LRG sample.}\label{fig:cmaghist}
\end{figure}

In Figures \ref{fig:cmag} and \ref{fig:cmaghist} we show the colour magnitude diagrams for those galaxies in our superclean sample which have spectroscopic magnitudes. The magnitudes and colours are based on absolute magnitudes calculated using \texttt{kcorrect v4\_1\_4} \citep{BlantonRoweis}. The elliptical galaxies in the sample have a mean $u$-$r$ of 2.55, significantly redder than the spirals (mean $u$-$r$=1.85).  These results correspond to the classic `red sequence' of early-type galaxies found by previous studies (e.g. \citet{Sandage2, Bower}), with the blue galaxies existing not on a tightly defined sequence but rather in a `blue cloud'. The division between the two is not straightforward, however. For example, close inspection of Figure \ref{fig:cmaghist} reveals that the sample contains populations of both blue elliptical galaxies (which are discussed in companion paper to this, Schawinski et al. 2008) and red spirals (the morphology-density relation for which is shown in Bamford et al. 2008 and which will be discussed in a future paper). 

In particular, Figure \ref{fig:cmaghist} includes a fit to the data with two Gaussians.  The rest frame colours used in these plots are calculated using k-corrections from \citet{BlantonRoweis}.  The combined result is reasonable, but as expected from the discussion above the two Gaussians do not clearly divide spiral from elliptical galaxies. In particular, the `blue elliptical' population forms a substantial contribution to the blue side of the redder of the two gaussians.  This result illustrates the importance of true morphological classification; even a sophisticated division between `red' and `blue' systems will not entirely separate the two morphological types.

\begin{figure}
\includegraphics[width=0.45\textwidth,angle=-90]{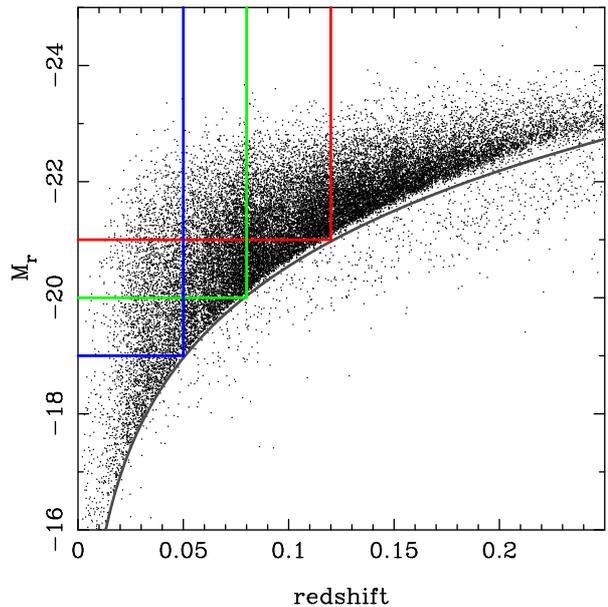}
\caption{Cuts applied to create volume limited subsamples from the clean sample. The curve shows the r=17.77 line converted to $M_r$ using the distance modulus but neglecting k-corrections, and corresponds to the main galaxy sample limit for an object with a flat spectrum. Points below (and some just above) this line are drawn from the LRG sample.}\label{fig:vollim}
\end{figure}

In order to explore further the properties of our sample in colour-magnitude space, we construct three volume-limited subsamples from those objects in the clean sample for which spectroscopic redshifts have been obtained. In order to improve confidence in the data, samples were constructed both for $r<$17.77 (solid lines) and $r<$17.0 (dashed lines). The cuts applied are illustrated in Figure \ref{fig:vollim}. The most luminous sample is dominated by elliptical galaxies, with a elliptical:spiral ratio of 1.99. The intermediate sample has a ratio of 0.98, and is thus evenly split between the two classes, whereas the sample including the faintest galaxies is dominated by spirals, with a ratio of 0.57.\label{sec:efrac}

\begin{figure}
\includegraphics[width=0.45\textwidth]{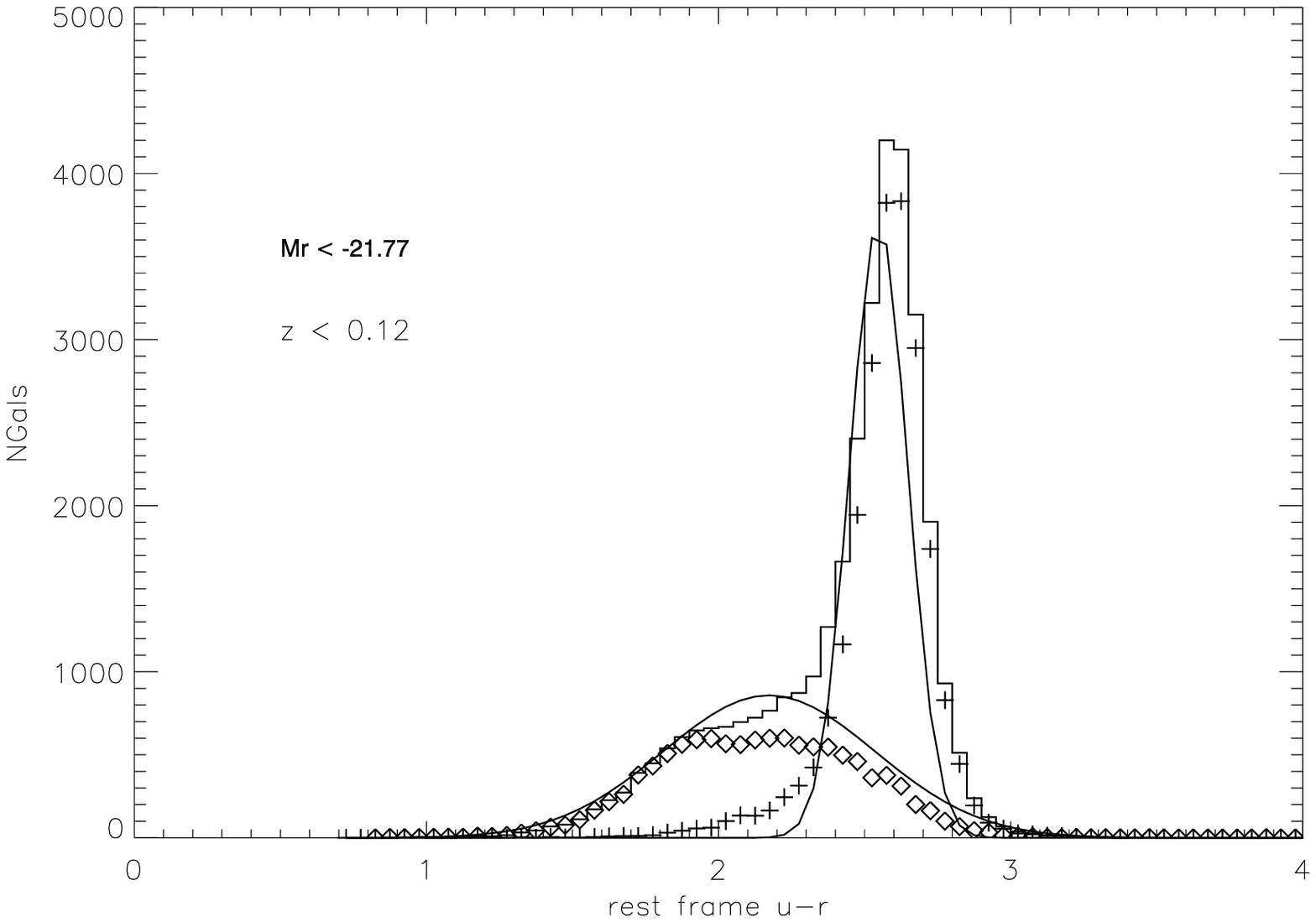}
\includegraphics[width=0.45\textwidth]{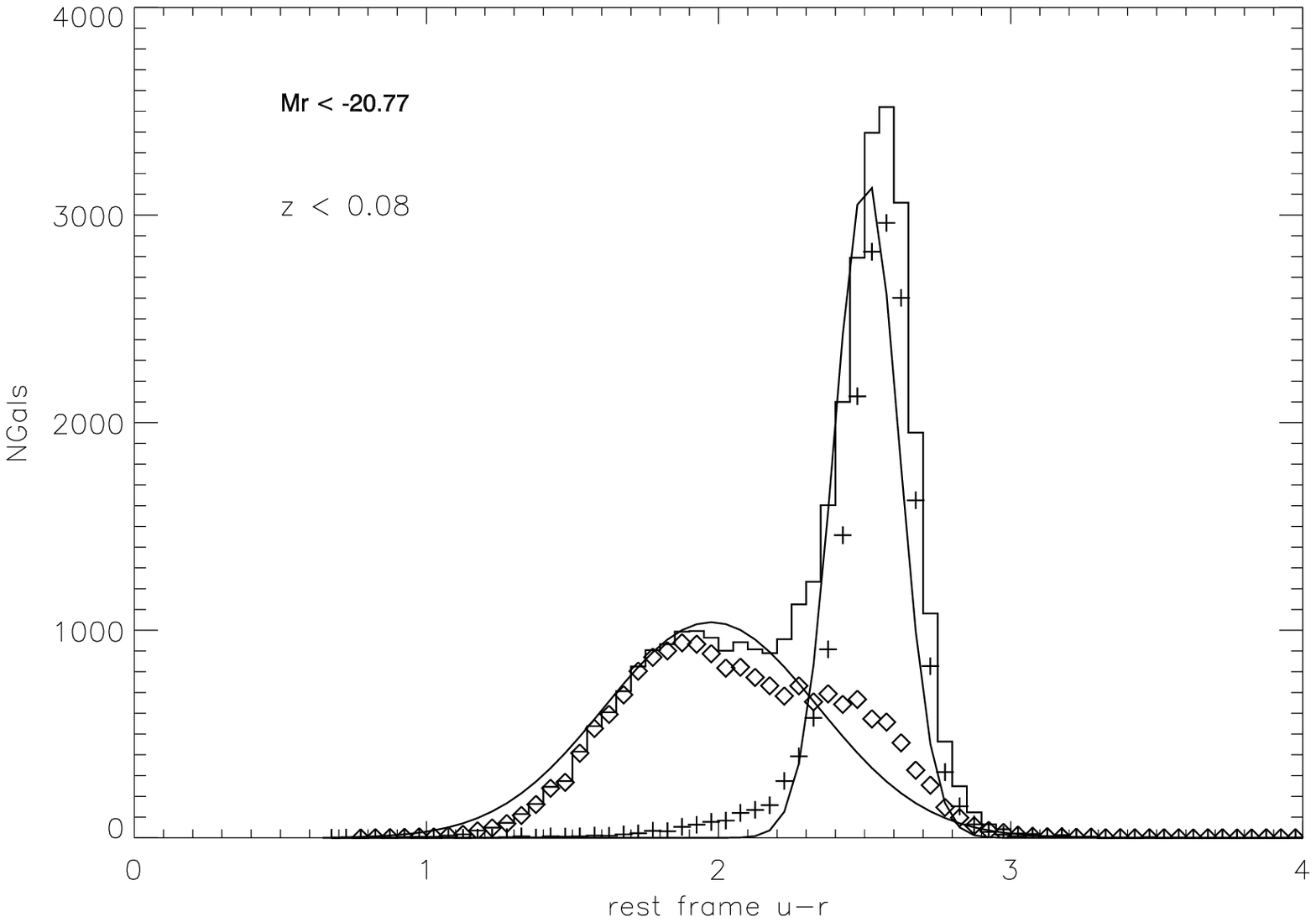}
\includegraphics[width=0.45\textwidth]{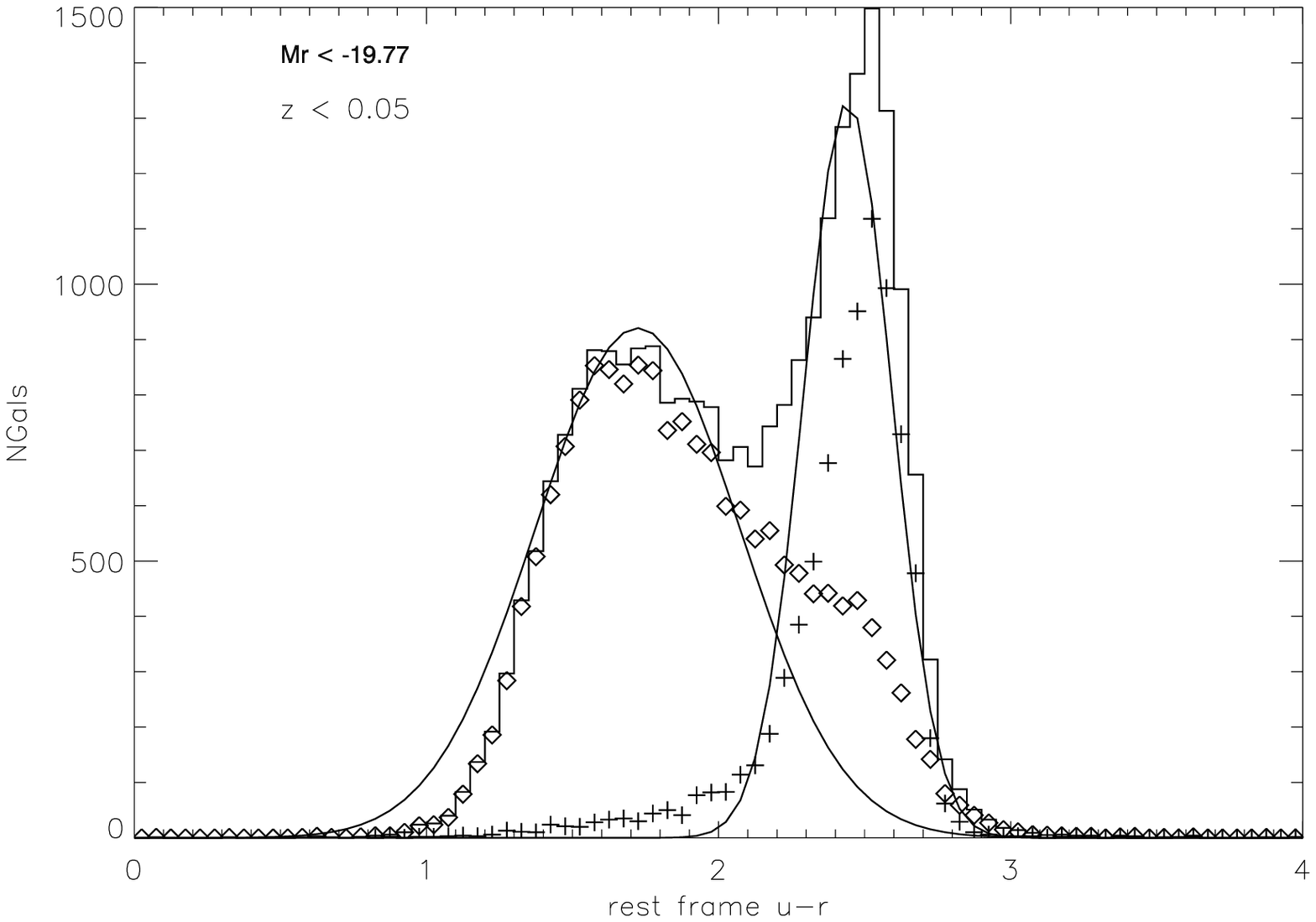}
\caption{Colour-magnitude diagrams for our set of three volume limited samples. Petrosian magnitudes are used, and k-corrections and absolute magnitudes derived from spectroscopic redshifts. As in Figure \ref{fig:cmaghist}, crosses represent ellipticals, diamonds spirals and the histogram the combined data set. Gaussian fits were made to the combined data (spirals and ellipticals) but only the individual Gaussians are shown. The vertical dotted lines are the limits proposed by \citep{Baldry} for dividing red and blue galaxies; systems to the right of this line are defined as red. As definied in equation \ref{eq:baldry}, this limit depends on the absolute magnitude and we thus show the limit for both minimum and maximum $M_r$ in each range.}\label{fig:baldry}
\end{figure}

Colour-magnitude diagrams for each of these subsamples are shown in Figure \ref{fig:baldry}. We also show Gaussian fits to the data based on those in \citet{Baldry}. Baldry et al. divide galaxies drawn from the SDSS into red and blue systems, defining a galaxy as red if $C_{ur}'>C_{ur}$ where $C_{ur}$ is the rest-frame (k-corrected) $u$-$r$ colour and 

\begin{equation}\label{eq:baldry}
C_{ur}'=2.06-0.244\tanh\left(\frac{M_r+20.07}{1.09}\right).
\end{equation}

Fits to the data shown in Figure \ref{fig:baldry} are gaussians with the same mean and variance as those derived in Baldry et al. These gaussians were then normalized to fit our data set. 

We show the results in Figure \ref{fig:baldry}, where some general trends are immediately apparent. The proportion of galaxies classified as ellipticals is larger in the sample which includes only the most luminous galaxies. The results also confirm as before that in none of the three cass the two distributions (red and blue galaxies) which would be derived in the absence of morphological information cannot be simply interpreted as corresponding to `early' and `late' type galaxies. It is not possible to define a single colour with which to divide the two classes of galaxy; rather the distributions overlap to a large extent. 

The biggest difference between the populations inferred from gaussian fitting and those obtained by our morphological classification is the presence of a substantial number of red galaxies which were classified as spiral systems in the lower luminosity samples. In fact, the population of galaxies which we classify as morphologically spiral contains a substantial number of systems with $u$-$r$ colours greater than $\sim 2.2$. Many of these systems may actually be lenticulars; however, distinguishing between well resolved edge-on S0 galaxies from an edge-on spiral is impossible by visual classification alone. Despite this contamination, however, true red spirals do exist in the data and morphological and colour bimodality are - at least for this intriguing population - decoupled. There is a corresponding population of blue elliptical galaxies, the most extreme examples of which are the subject of a companion paper (Schawinski et al. 2008), but they are less significant here. 

\section{Conclusion}

We have described Galaxy Zoo, a web-based project which invited the public to classify galaxies imaged as part of the Sloan Digital Sky Survey. By combining the classifications of more than 100,000 participants in the largest astronomical collaboration in history, we are able to produce catalogues of galaxy morphology which agree with those compiled by professional astronomers to an accuracy of better than 10\%. Our results thus suggest that the general public can reliably classify large sets of galaxies with a similar accuracy as can professional astronomers. The largest of the Galaxy Zoo catalogues includes more than 300,000 galaxies reliably classified at more than 5$\sigma$ confidence according to morphology, a factor of ten larger than previous work. Due to the repeated, independent classifications of the same object it is possible to quantify the errors in the classification, and produce catalogues of differing fidelity for different purposes (such as the clean and superclean catalogues discussed here). By examining a volume-limited subset of the data in colour-magnitude space we illustrate the differences between the colour and morphological bimodalities in the data. The presence of a substantial number of red galaxies classified as spiral in particular underlines the importance of morphological classification; our results show that a traditional morphological classification cannot be reproduced by cuts on colour alone. 

\section*{Acknowledgments}
In addition to the contribution from Galaxy Zoo volunteers, we also acknowledge invaluable contributions from Edd Edmondson, Pete Wilton, Alice Sheppard and Danny Locksmith. We thank Prof. Joe Silk for his encouragement. CJL acknowledges support from the STFC Science in Society Program.

Funding for the SDSS and SDSS-II has been provided by the Alfred P. Sloan Foundation, the Participating Institutions, the National Science Foundation, the U.S. Department of Energy, the National Aeronautics and Space Administration, the Japanese Monbukagakusho, the Max Planck Society, and the Higher Education Funding Council for England. The SDSS Web Site is http://www.sdss.org/.

The SDSS is managed by the Astrophysical Research Consortium for the Participating Institutions. The Participating Institutions are the American Museum of Natural History, Astrophysical Institute Potsdam, University of Basel, University of Cambridge, Case Western Reserve University, University of Chicago, Drexel University, Fermilab, the Institute for Advanced Study, the Japan Participation Group, Johns Hopkins University, the Joint Institute for Nuclear Astrophysics, the Kavli Institute for Particle Astrophysics and Cosmology, the Korean Scientist Group, the Chinese Academy of Sciences (LAMOST), Los Alamos National Laboratory, the Max-Planck-Institute for Astronomy (MPIA), the Max-Planck-Institute for Astrophysics (MPA), New Mexico State University, Ohio State University, University of Pittsburgh, University of Portsmouth, Princeton University, the United States Naval Observatory, and the University of Washington.

\label{lastpage}

\end{document}